\documentclass[10pt, superscriptaddress, reprint, amsmath, amssymb, aps, pra, showpacs]{revtex4-2}

\usepackage{amsmath,amssymb,mathrsfs}
\usepackage{cases}
\usepackage{graphicx}
\usepackage{dcolumn}
\usepackage{bm}
\usepackage{color}
\usepackage{amsfonts,latexsym}
\usepackage{enumerate}
\usepackage{rotating}
\usepackage{epstopdf}
\usepackage{braket}
\usepackage{hyperref}
\hypersetup{
	colorlinks=true,
	linkcolor=blue,
	filecolor=blue,
	urlcolor=blue,
	citecolor=magenta
}

\newcommand{\Eq}[1]{Eq.~\eqref{#1}}
\newcommand{\eq}[1]{\eqref{#1}}
\newcommand{\Fig}[1]{Fig.~\ref{#1}}

\newcommand{\beq}{\begin{equation}}
	\newcommand{\eeq}{\end{equation}}
\newcommand{\beqa}{\begin{eqnarray}}
	\newcommand{\eeqa}{\end{eqnarray}}
\newcommand{\Beqa}{\begin{eqnarray*}}
	\newcommand{\Eeqa}{\end{eqnarray*}}

\def\bal#1\eal{\begin{align}#1\end{align}}
\def\Bal#1\Eal{\begin{align*}#1\end{align*}}

\newcommand{\nn}{\nonumber}

\newcommand{\me}{\mathrm{e}}

\newcommand{\mi}{\mathrm{i}}

\newcommand{\dif}{\mathrm{d}}

\newcommand{\mre}{\mathrm{Re}}	
\newcommand{\mim}{\mathrm{Im}}	
\newcommand{\Li}{\mathrm{Li}} 	

\newcommand{\vect}[1]{\mathbf{#1}}
\newcommand{\overbar}[1]{\mkern 1.5mu\overline{\mkern-1.5mu#1\mkern-1.5mu}\mkern 1.5mu}

\begin{document}
	
	\title{Three-body recombination in a single-component Fermi gas with $p$-wave interaction}
	
	\author{Shangguo Zhu}
	\email[]{zhushangguo@ioe.ac.cn}
	\affiliation{Research Center on Vector Optical Fields, Institute of Optics and Electronics, Chinese Academy of Sciences, Chengdu 610209, China} 
	\affiliation{Department of Physics and HKU-UCAS Joint Institute for Theoretical and Computational Physics at Hong Kong, Guangdong-Hong Kong Joint Laboratory of Quantum Matter, University of Hong Kong, Hong Kong, China}

	\author{Zhenhua Yu}
	\email[]{huazhenyu2000@gmail.com}
	\affiliation{Guangdong Provincial Key Laboratory of Quantum Metrology and Sensing, and School of Physics and Astronomy, Sun Yat-Sen University, Zhuhai 519082, China}
	\affiliation{State Key Laboratory of Optoelectronic Materials and Technologies, Sun Yat-Sen University, Guangzhou 510275, China}
	
	\author{Shizhong Zhang}
	\email[]{shizhong@hku.hk}
	\affiliation{Department of Physics and HKU-UCAS Joint Institute for Theoretical and Computational Physics at Hong Kong, Guangdong-Hong Kong Joint Laboratory of Quantum Matter, University of Hong Kong, Hong Kong, China}
	\date{\today}
	
	\begin{abstract} 
		We study the three-body recombination of identical fermionic atoms. Using a zero-range model for the $p$-wave interaction, we show that the rate constant of three-body recombination into weakly bound $p$-wave dimers can be written as $\alpha_{\rm rec} \propto v^{5/2}R^{1/2} k_T^4 (1+ C k_T^2 l_{\rm d}^2)$ for large and positive scattering volume $v$. Here $R$ is the $p$-wave effective range, $k_T^2$ gives the average thermal kinetic energy of the colliding atoms, and $l_{\rm d}$ is the size of the $p$-wave dimer. The leading term is different from the usually stated $v^{8/3}$-scaling law, but is consistent with an earlier two-channel calculation. For the subleading term, we compute the constant $C$ by solving the relevant three-body problem perturbatively when the parameter $\gamma\equiv R/v^{1/3}$ is small. The additional $C k_T^2 l_{\rm d}^2$ term provides important corrections for the temperature and interaction dependence of $\alpha_{\rm rec}$, especially close to resonance when $k_T l_{\rm d}$ is relatively large. 
	\end{abstract}
	\maketitle

	\section{Introduction}
	
	Three-body recombination in ultracold gases often acts as an obstacle that limits the lifetime of trapped gases. On the other hand, it can also serve as an excellent probe for observing remarkable few-body phenomena~\cite{Esry1999, Kraemer2006, Hammer2007, Stecher2009, Ferlaino2009, Efimov2009a, Efimov2009b, Efimov2017, Efimov2020}. During the recombination process, three atoms collide and two of them form a dimer (a molecular bound state); the released binding energy allows the dimer and the third atom to escape from the trap and causes atom loss. It turned out that the three-body processes play a much more prominent role in a single-component Fermi gas close to $p$-wave Feshbach resonance, and the related stability problem has been extensively studied~\cite{Levinsen2007, Levinsen2008, Cooper2009, Han2009, Zinner2010, Waseem2017, Kurlov2017, Zhou2017, Pan2018, Hu2018, Jiang2018, Chang2020}. For a comprehensive understanding, it is vital to investigate microscopically the three-body recombination of identical fermionic atoms.

	For identical fermionic atoms, the recombination rate constant $\alpha_{\rm rec}$ contributes to the inelastic part of the three-body scattering hypervolume~\cite{Tan2008,Zhu2017,Mestrom2019,Wang2021} and scales as $E^2$ when the collision energy $E$ is small according to the threshold law~\cite{Esry2001}. From dimensional analysis, $\alpha_{\rm rec}$ is proportional to a quantity with dimension $({\rm Length})^8$ and should be determined by the low-energy scattering parameters, the scattering volume $v$, and the effective range $R$, defined in the effective range expansion 
	\begin{equation}
		p^3 \cot \delta_1 = -\frac{1}{v} - \frac{1}{R}p^2+\cdots, \label{def_effectiverange}
	\end{equation}
	with $p$ the wave vector for the relative motion and $\delta_1$ the $p$-wave phase shift. Here the ellipsis represents higher-order terms that are not relevant for our discussion below.

	The numerical calculation with an adiabatic hyperspherical approach predicted that $\alpha_{\rm rec}\propto |v|^{8/3}$ for $R / |v|^{1/3} \gtrsim 0.1$ (irrespective of the sign of $v$)~\cite{Suno2003}. 
	For negative and small $v$ where recombination leads to the formation of deep dimers, $\alpha_{\rm rec}$ was found to be consistent with $|v|^{8/3}$ scaling based on the numerically solved Faddeev integral equation~\cite{Schmidt2020} and is also consistent with the recent experiments~\cite{Yoshida2018, Waseem2018, Waseem2019} in the same parameter regime. 
	However, an analytic calculation based on a two-channel model pointed out that $\alpha_{\rm rec} \propto v^{5/2}R^{1/2}$ for recombinations into shallow dimers at $v>0$~\cite{Jona-Lasinio2008}. Although the power dependence on $v$ that can be checked in actual experiments differs only by  $6\%$, it is important to establish the correct dependence theoretically. In addition, experiments~\cite{Yoshida2018, Waseem2019} have shown clearly the deviation from the leading scaling with $v$ close to the strong interacting regime. It is thus also necessary to establish the corrections to the leading dependences in order to quantitatively compare with experiments.

	In this work we study analytically the recombination of three identical fermionic atoms into a shallow $p$-wave dimer at large and positive $v$.  
	Using a zero-range model for the $p$-wave interaction, we obtain an ansatz for the three-body wave function and derive the integro-differential equation for the atom-dimer motion. 
	Based on the smallness of $\gamma \equiv R / v^{1/3}$, we first confirm the leading $v^{5/2}R^{1/2}$ scaling law as obtained by the two-channel model~\cite{Jona-Lasinio2008}, different from the $v^{8/3}$ dependence suggested in Ref.~\cite{Suno2003}. Furthermore, we obtain corrections to the leading scaling and show that it is characterized by the parameter $k_T^2 l^2_{\rm d}$, where $k_T^2$ gives the average thermal kinetic energy of the colliding atoms and $l_{\rm d} \equiv \sqrt{v/R}$ is the size of the shallow dimer.

	\section{Zero-range model}
	Let us start our discussion with the case of two interacting identical fermions. Outside the range of the short-range interactions, the relative wave function takes the form 
	\begin{equation}
		\psi_{\rm 2b} (\vect{r})=\sum_{\mu=1}^3 \mathcal{N}_\mu [ j_1(pr) \cot\delta_1 - y_1(pr)] Y^{(\mu)}(\hat{\bf r}),
	\end{equation}
	where $j_1(pr)$ and $y_1(pr)$ are the spherical Bessel functions of the first and the second kind, respectively, $Y^{(\mu)}(\hat{\vect{r}}) \equiv \sqrt{{3}/(4\pi)}{r^{(\mu)}}/{r} $ is the $p$-wave real spherical harmonic, the superscript $\mu$ represents the $\mu$th component of a vector, and $\mathcal{N}_\mu$ are the coefficients.
	Given the $p$-wave phase shift characterized by both $v$ and $R$ in Eq.~(\ref{def_effectiverange}), $\psi_{\rm 2b}(\vect{r})$ behaves at short distance as~\cite{Pricoupenko2006,Peng2014}
	\begin{equation}
		\label{bc}
		\psi_{\rm 2b} (\vect{r}) = \sum_{\mu=1}^{3}\left[\frac{\mathcal{A}_{\mu} }{r^2}+\mathcal{B}_{\mu} +\mathcal{C}_{\mu} r \right] Y^{(\mu)}(\hat{\vect{r}}) +O(r^2),
	\end{equation}
	with coefficients satisfying the constraint
	\begin{equation}
		\label{constraint}
		v^{-1} \mathcal{A}_{\mu}+ 2R^{-1} \mathcal{B}_{\mu}  +3 \mathcal{C}_{\mu} =0.
	\end{equation}
	The above expansion with the constraint can be viewed as the short-distance boundary condition for the $p$-wave channel. 
	This is equivalent to treating the interaction term as a zero-range pseudopotential~\cite{Pricoupenko2006}. In fact, if we were to extend the wave function to zero distance, we can show that
	\begin{equation}
		(\nabla^2+p^2)\psi_{\rm 2b}(\vect{r})=\sum_{\mu=1}^{3}\frac{\sqrt{12\pi} }{p^2} \mathcal{N}_\mu \frac{\partial \delta(\vect{r} )}{\partial r^{(\mu)}}, 
	\end{equation}
	where the right-hand side acts as the pseudopotential for the $p$-wave scattering.

	Now let us turn to the three identical fermions. As usual, let us define the Jacobi vectors for the three-body problem
	\begin{align}
		\vect{x}_i &= \vect{r}_{j} - \vect{r}_{k} \\
		\vect{y}_{i} &= \frac{2}{\sqrt{3}}\left[\vect{r}_{i} - \frac{\vect{r}_{j} + \vect{r}_{k}}{2}\right]
	\end{align}
	with $(ijk)=(123)$, $(231)$, or $(312)$, where $(\vect{r}_1, \vect{r}_2, \vect{r}_3)$ are the position vectors of the three atoms. In most cases, we can suppress the subscript and let $\vect{x}\equiv \vect{x}_1$ and $\vect{y}\equiv \vect{y}_1$. The other pairs of Jacobi vectors can be expressed as linear combinations of $\vect{x}$ and $\vect{y}$,
	\begin{align}
		\vect{x}_2 &= -\frac{1}{2}\vect{x}-\frac{\sqrt{3}}{2}\vect{y}, \qquad \vect{y}_2= \frac{\sqrt{3}}{2}\vect{x}-\frac{1}{2}\vect{y},\\
		\vect{x}_3 &= -\frac{1}{2}\vect{x}+\frac{\sqrt{3}}{2}\vect{y}, \qquad \vect{y}_3= -\frac{\sqrt{3}}{2}\vect{x}-\frac{1}{2}\vect{y}.
	\end{align}
	
	We use units such that the reduced Planck constant $\hbar = 1$ and the mass of the fermionic atom $M=1$. In the limit when $|\vect{x}|\to 0$ while $|\vect{y}|$ remains finite, a suitable ansatz for the three-body wave function would be to replace $\mathcal{N}_\mu$ by a function of $\vect{y}$, $(p^2/4\pi)f_\mu({\bf y})$. Thus we obtain the Schr\"odinger equation for the three-body wave function $\psi(\vect{x},\vect{y})$ in the center-of-mass frame
	\begin{equation}
		\label{scheqcom}
		\left(\nabla_{\vect{x}}^{2} + \nabla_{\vect{y}}^{2} + E\right)\psi = \sqrt{\frac{3}{4\pi}} \sum_{i,\mu=1}^{3}f_{\mu}(\vect{y}_{i}) \frac{\partial \delta( \vect{x}_{i} )}{\partial x_{i}^{(\mu)}},
	\end{equation}
	where $E$ gives the collision energy.
	The unknown function $f_{\mu}(\vect{y})$ describes the atom-dimer motion and is to be determined by imposing the boundary conditions as in the two-body case \eq{bc} except that now the coefficients $\mathcal{A}_{\mu}, \mathcal{B}_{\mu}$, and $\mathcal{C}_{\mu}$ are all functions characterized by $f_\mu(\vect{y})$.

	\section{Equation for the atom-dimer motion}
	
	The three-body wave function, satisfying \Eq{scheqcom}, can be expressed in the general form
	\begin{equation}
		\label{trialpsi}
		\psi =  \psi_0 + \sum_{i,\mu=1}^{3} \int \dif^3 \vect{y}' f_{\mu}(\vect{y}') G_{\mu}(\vect{x}_{i},\vect{y}_{i}-\vect{y}' ), 
	\end{equation}
	where 
	\begin{equation}
		G_{\mu}(\vect{x}, \vect{y}) = \frac{\kappa^3}{8 \pi^3} \frac{K_{3} (\kappa \sqrt{x^2+y^2}) }{(x^2+y^2)^{3/2}} x Y^{(\mu)}(\vect{x})
	\end{equation} 
	is the derivative with respect to $x^{(\mu)}$ of the Green's function for the differential operator in Eq.~(\ref{scheqcom}). Here $K_{n}(r)$ is the exponentially decaying Bessel function. 
	We use the convention that $\kappa = - \mi \sqrt{E}$ if $E>0$ or $\sqrt{-E}$ if $E<0$. In addition, $\psi_0$ is the solution to the homogeneous differential equation (\ref{scheqcom}). Its explicit form is the fully antisymmetrized wave function of three free fermionic atoms and is given by 
	\begin{equation}
		\psi_0 = \sqrt{\frac{1}{6}} V^{-3/2} \sum_{P} (-1)^{P} P \exp \sum_{i=1}^{3} \mi \overbar{\vect{b}}_i \cdot \vect{r}_i.
	\end{equation}
	Here $P$ represents the permutation of the three atoms, $V$ is the system volume, and $\overbar{\vect{b}}_i$ is the wave vector of the individual atom and the collision energy $E= (\overbar{b}_1^2 + \overbar{b}_2^2 +\overbar{b}_3^2 )/2$. The $\psi_0$ term is absent when $E<0$. As we consider the recombination problem, we require $E>0$. 
	
	The wave function $\psi$ satisfies the boundary condition \Eq{bc} at short interparticle distances that requires us to study the small-$x$ behavior of $\psi$. We can rewrite $\psi$ as a sum of four terms $\psi = I_1 + I_2 + I_3 + I_4$, 
	\begin{align} 
		I_1 &= \psi_0,\\ 
		I_2 &= \sum_{\mu=1}^{3} \int \dif^3 y'~ f_{\mu}(\vect{y})G_{\mu}(\vect{x},\vect{y}-\vect{y}' ),\\
		I_3 &= \sum_{\mu=1}^{3} \int \dif^3 y'~ [ f_{\mu}(\vect{y}')-f_{\mu}(\vect{y}) ] G_{\mu}(\vect{x},\vect{y}-\vect{y}' ),\\
		I_4 &= \sum_{\mu=1}^{3} \int \dif^3 y'~ [G_{\mu}(\vect{x}_2 , \vect{y}_2-\vect{y}' )+G_{\mu}(\vect{x}_3 , \vect{y}_3-\vect{y}' )].
	\end{align} 	
	The first term $I_1=\psi_0$ can be written equally as 
	\begin{equation}
		I_1 = 2\mi \sqrt{1/6}V^{-3/2} \sum_{i=1}^{3}  \sin(\vect{b}_{i}\cdot\vect{x})\exp(\mi \vect{B}_{i}\cdot\vect{y})
	\end{equation} 
	where $\vect{b}_i = (\overbar{\vect{b}}_j - \overbar{\vect{b}}_k) /2$ and $\vect{B}_i = [ \overbar{\vect{b}}_i - ( \overbar{\vect{b}}_j + \overbar{\vect{b}}_k ) / 2  ] / \sqrt{3}$ with $(ijk)=(123)$, $(231)$, or $(312)$. For small $x$, it can be expanded as 
	\begin{equation}
		I_1 = \psi_0 = \sum_{\mu=1}^{3}C_{\mu}(\vect{y}) x Y^{(\mu)}(\hat{\vect{x}}) +O(x^3),
	\end{equation}
	where the explicit form for $C_{\mu}(\vect{y})$ is given by 
	\begin{equation}
		C_{\mu}(\vect{y}) = \frac{2}{3} \mi \sqrt{2\pi} V^{-3/2} \sum_{\mu=1}^{3} b_{i}^{(\mu)} \exp(\mi \vect{B}_{i}\cdot\vect{y}).
	\end{equation}	
	For $I_2$ we note that $\int \dif^3 y~ G^{(\mu)}(\vect{x},\vect{y})  = \frac{1+\kappa x}{4\pi x^2}\me^{-\kappa x}Y^{(\mu)}(\hat{\vect{x}})$ and $I_2$ can be expanded as
	\begin{equation}
		I_2 = \sum_{\mu=1}^{3} \Big[\frac{1}{4\pi x^2} -\frac{\kappa^2}{8\pi} +\frac{\kappa^3 x}{12\pi}  +O(x^2) \Big] Y^{(\mu)}(\hat{\vect{x}}) f_{\mu}(\vect{y}).
	\end{equation}	
	The third term can be expanded as
	\begin{align}
		I_3 &=  \sum_{\mu=1}^{3} \Big\{ \frac{1}{8\pi} \nabla_{\vect{y}}^2 f_{\mu}(\vect{y})  \nn\\ 
		& + \frac{\kappa^3 x}{8\pi^3}  \mathcal{P} \int \dif^3 y'~  \Big[f_{\mu}(\vect{y}')-f_{\mu}(\vect{y})\Big] \frac{K_3( \kappa |\vect{y}-\vect{y}'| )}{|\vect{y}-\vect{y}'|^3}  \nn\\
		&+ O(x^2)\Big\} Y^{(\mu)}(\hat{\vect{x}}),
	\end{align}
	where $\mathcal{P}$ denotes the generalized principal value integral~\cite{Fox1957}. 
	For an integrand with a pole of order $n+1$, the principal value integral is defined as
	\begin{align}
		\mathcal{P} \int_{a_1}^{a_2} &  \frac{h(x)}{(x-x_0)^{n+1}} \dif x  \equiv \lim_{\epsilon\to 0^+} \Big[ \int_{a_1}^{x_0-\epsilon}  \frac{h(x)}{(x-x_0)^{n+1}} \dif x  \nn\\
		& +\int_{x_0+\epsilon}^{a_2}  \frac{h(x)}{(x-x_0)^{n+1}} \dif x  - \mathcal{E}_{n}(x_0,\epsilon)\Big],
	\end{align}
	where $a_1 < x_0 < a_2$ and $h(x)$ is a smooth function at $a_1 < x < a_2$.  Here $\mathcal{E}_{0}(x_0,\epsilon)=0$, $\mathcal{E}_{1}(x_0,\epsilon)=\frac{2}{\epsilon} h(x_0)$,
	$\mathcal{E}_{2}(x_0,\epsilon) = \frac{2}{\epsilon} h'(x_0)$, and $\mathcal{E}_{3}(x_0,\epsilon) = \frac{2h(x_0)}{3\epsilon^3}  +\frac{h''(x_0)}{\epsilon}$. 
	For a general $n\ge 1$, $\mathcal{E}_{n}(x_0,\epsilon) = \sum_{k=0}^{n-1} \frac{h^{(k)}(x_0)}{k!(n-k)} \frac{1-(-1)^{n-k}}{\epsilon^{n-k}}$. 
	For $n=0$ and $n=1$, $\mathcal{P}$ represents the Cauchy principal value and the Hadamard finite part, respectively. 
	
	The fourth term has the small-$x$ expansion
	\begin{align}
		I_4 &=- \frac{\kappa^6 x}{8\pi^3} \sum_{\mu=1}^{3} \int \dif^3 y'~ \Big\{  g_3\Big(\kappa \chi \Big)  f_{\mu}(\vect{y}')  \nn\\
		&+  \sqrt{3 \pi} \kappa^2 g_4\Big(\kappa  \chi \Big)  \Big[\vect{y} \cdot \vect{f}(\vect{y}') \Big]  y' Y^{(\mu)}(\hat{\vect{y}}')\Big\} Y^{(\mu)}(\hat{\vect{x}}) \nn\\
		& \quad +O(x^3).
	\end{align}
	Here and in the following discussion we use the shorthand $g_n(x) \equiv \frac{K_n(x)}{x^n}$, $\xi \equiv |\vect{y}' - \vect{y}|$ and $\chi \equiv \sqrt{y^2+y'^2+\vect{y}\cdot\vect{y}'}$.

	Now we can extract the three coefficients $\mathcal{A}_{\mu}$, $\mathcal{B}_{\mu}$, and $\mathcal{C}_{\mu}$ in the expansion $\psi = \sum_{\mu=1}^{3}[ \mathcal{A}_{\mu} x^{-2} +\mathcal{B}_{\mu}+\mathcal{C}_{\mu}x +O(x^2)] Y^{(\mu)}(\hat{\vect{x}}) +O(x^3)$. 
	From the constraint \eq{constraint} we obtain the integro-differential equation for the atom-dimer function $f_{\mu}(\vect{y})$,
	\begin{equation}
		\label{eqf}
		12\pi C_{\mu}(\vect{y}) = \Big[-\frac{1}{v} +\frac{\kappa^2-\nabla_{\vect{y}}^{2}}{R}-\kappa^3 + \hat{L}\Big] f_{\mu}(\vect{y}) + \hat{S}_\mu \vect{f}(\vect{y}),
	\end{equation}
	where $\hat{L}$ and $\hat{S}_{\mu}$ are integral operators
	\begin{align}
		\hat{L} f(\vect{y}) &= \frac{3 \kappa^3}{2\pi^2} \mathcal{P} \int \dif^3 y'~  \Big\{ \Big[f(\vect{y})-f(\vect{y}') \Big] \frac{K_3(\kappa \xi )}{\xi^3}  \nn\\  
		& \qquad\qquad\qquad\qquad + f(\vect{y}') \frac{K_3(\kappa \chi )}{\chi^3} \Big\},\label{Loperator} \\  
		\hat{S}_{\mu} \vect{f}(\vect{y}) &=  \frac{3\sqrt{3\pi} \kappa^4}{2\pi^2}  \int \dif^3 y'~  \frac{K_4(\kappa \chi )}{\chi^4} \Big[ \vect{y} \cdot \vect{f}(\vect{y}')  \Big] y' Y^{(\mu)}(\hat{\vect{y}}')\label{Soperator}
	\end{align}
	and $\vect{f} \equiv (f_{1}, f_{2}, f_{3})$. 
	Note that $\hat{L}$ conserves angular momentum, while $\hat{S}_{\mu}$ does not. 
	This integro-differential equation is akin to the Skorniakov--Ter-Martirosian integral equation derived for three bosons~\cite{Skorniakov1956,Petrov2004}.

	\section{Leading order at low energies}
	
	According to the threshold law~\cite{Esry2001}, the $p$-wave channel of the atom-dimer motion gives the dominant contribution (proportional to $E^2$) to the recombination rate at low energies. For the leading-order approximation, we only consider the $p$-wave channel, and express the solution as 
	\begin{equation}
		f_\mu(\vect{y}) = \sum_{\nu=1}^{3} \mathcal{F}_{\mu\nu}(y) Y^{(\nu)}(\hat{\vect{y}})\label{exp_fmu},
	\end{equation} 
	where $\mathcal{F}_{\mu\nu}(y)$ can be expanded as 
	\begin{equation}
		\label{Fexpansion}
		\mathcal{F}_{\mu\nu}(y) = f_{\mu\nu}(y) + \kappa^2 h_{\mu\nu}(y) +\cdots
	\end{equation}
	at low energies. 
	Note that in \Eq{eqf} the $p$-wave component of the inhomogeneous term is given by 
	\begin{equation}
		\label{cpwave}
		C_{\mu\nu}^{l=1}(y) = \int \dif \Omega_{y} Y^{(\nu)}(\hat{\vect{y}}) C_{\mu}(\vect{y}) =  D_{\mu\nu}^{0} y + D_{\mu\nu}^{1} y^3 + O(\kappa^6),
	\end{equation}
	where $D_{\mu\nu}^{0} = -({2\sqrt{6} \pi }/{3})V^{-3/2} \sum_{\rho=1}^{3}\epsilon_{\rho\mu\nu} (\vect{b}\times \vect{B} )^{(\rho)} $, of order $\kappa^2$, and $D_{\mu \nu}^{1} = (2\sqrt{6}\pi /45)  V^{-3/2} \sum_{i=1}^{3} b_{i}^{(\mu)} B_{i}^{(\nu)} B_{i}^{2}$, of order $\kappa^4$. 
	Here $\epsilon_{\rho\mu\nu}$ is the Levi-Civit\`{a} symbol and $\vect{b} \equiv \vect{b}_1  $ and $\vect{B}= \vect{B}_1 $ are the wave vectors associated with $\vect{x}$ and $\vect{y}$, respectively. 
	Since $D_{\mu\nu}^{0} =  - D_{\nu\mu}^{0}$, we find $f_{\mu\nu}(y) = - f_{\nu\mu}(y)$ from \Eq{eqf}.

	On the other hand, $\hat{S}_{\mu}$ does not conserve the angular momentum and gives an extra $f$-wave part when acting on the $p$-wave channel. In the leading-order approximation of \Eq{eqf}, we neglect the $f$-wave part and obtain (see Appendix~\ref{sect:operators}) 
	\begin{equation}
		\label{eqpwave}
		12\pi D_{\mu\nu}^{0} y = \left(-\frac{1}{v} + \frac{1}{R} \hat{K}_1 + \hat{A}_0 \right) f_{\mu\nu}(y),
	\end{equation}
	where
	\begin{equation}
		\hat{K}_l \equiv -  \frac{1}{y^2} \frac{\dif}{\dif y} y^2 \frac{\dif}{\dif y} + \frac{l(l+1)}{y^2} 
	\end{equation} 
	is the kinetic operator of angular momentum $l$ and $\hat{A}_0$ is an integral operator derived from the low-energy expansion of $\hat{L}$ and $\hat{S}_{\mu}$ and is given as
	\begin{align}
		\hat{A}_0 f(y) &=  -\frac{3}{\pi }  \mathcal{P} \int_0^\infty \dif y'~ f(y') \sum_{\sigma=\pm}\Big( \frac{2y'}{y \xi_{\sigma}^{4}}   +  \frac{\sigma}{y^2 \xi_{\sigma}^{2}}  \nn\\
		&\qquad\qquad\qquad +  \frac{16y'}{y \chi_{\sigma}^{4} } + \frac{16 \sigma }{y^2 \chi_{\sigma}^{2}}  \Big), 
	\end{align}
	where we used the shorthand $\xi_\pm \equiv |y\pm y'|$ and $\chi_\pm \equiv \sqrt{y^2+y'^2 \pm y y'} $. For $-4< \mre(s) <4$ we find $\hat{A}_0 y^{-s} = A_0(s) y^{-s-3}$, where
	\begin{align}	
		A_0(s) &= \frac{\sin\frac{\pi s}{6}}{\sin\frac{\pi s}{2}} \Big[ \frac{64}{\sqrt{3}} + s\big(s^2-36\big)\cot\frac{\pi s}{6} \nn\\
		&\qquad\qquad -2s \big(s^2-4\big) \sin\frac{\pi s}{3} \Big] . 
	\end{align}
	For simplicity, we further obtain the dimensionless form of \Eq{eqpwave},
	\begin{equation}
		\label{eqlambda}
		\lambda^{-5} = \Big( -1  + \hat{\mathbf{K}}_{1} + \gamma^{3/2}  \hat{\mathbf{A}}_0  \Big) F(\lambda),
	\end{equation}
	where $\gamma = R/v^{1/3}$ and $\lambda = y/l_{\rm d}$, with $l_{\rm d} = \sqrt{v/R} $ approximately the dimer size. In addition, $\hat{\mathbf{K}}_l \equiv l_{\rm d}^{2} \hat{K}_l$ and $\hat{\mathbf{A}}_0  \equiv l_{\rm d}^{3} \hat{A}_0 $. 
	We have defined the auxiliary function $F(\lambda)$ via
	\begin{align}
		\label{auxf}
		f_{\mu\nu}(y) &= 12\pi D_{\mu\nu}^{0} v l_{\rm d} \Big[- \lambda -  A_0(-1) \frac{\gamma^{3/2}}{\lambda^2} \nn\\
		&\qquad\qquad\qquad + \gamma^3  A_0(-1)  A_0(2) F(\lambda) \Big]. 
	\end{align}
	When approaching a $p$-wave Feshbach resonance, $\gamma$ is small. Equation \eq{eqlambda} can be solved perturbatively based on the smallness of $\gamma$. The general solution is a combination of a particular solution and the solutions to the homogeneous part of \Eq{eqlambda} and is then uniquely fixed by two proper boundary conditions, one at small $y$ and one at large $y$.

	Consider first the boundary condition at small $y$. 
	Note that the zero-range model of the $p$-wave interaction breaks down when we extend the wave function from outside the range of interaction into the inside~\cite{Pricoupenko2006,Braaten2012,Nishida2012}. 
	Thus, we need to impose a proper boundary condition whenever any two atoms reach a distance roughly the range of interaction $r_e$. 
	For simplicity, we impose the Dirichlet boundary $f_{\mu}(\vect{y}) = 0$ at $y = y_c$ with $y_c \gtrsim r_e$. 
	Note that $r_e$ is usually of order $R$.  
	This choice of boundary condition also avoids unphysical solutions. 
	We note that the function $A_0(s)$ has four complex zeros at $s\approx \pm (2.086\pm 1.103 \mi)$. 
	Then the solution behaves as $ F(\lambda) \sim \lambda^{2.086} \sin( 1.103 \ln \lambda + \tilde{\theta})$ with a phase $\tilde{\theta}$ at $\lambda\to 0$ (see Appendix~\ref{sect:smalllambda}). 
	The log-periodic feature indicates discrete scaling symmetry and suggests the existence of the Efimov effect~\cite{Efimov1970,Efimov1970a,Braaten2006,Braaten2007,Naidon2017}. 
	However, it has been proven that the Efimov effect is impossible for the $p$-wave interaction in three dimensions because it leads to the unphysical situation that the normalized wave function exhibits negative probability inside the range of interaction~\cite{Nishida2011,Braaten2012,Nishida2012}. 
	Moreover, from a Born-Oppenheimer analysis, the binding energy of the three-body system has an inverse-cubic dependence on the distance and this suggests the absence of scaling symmetry~\cite{Zhu2013,Efremov2013}.

	For the recombination problem, $f_{\mu}(\vect{y})$ should only contain the outgoing wave at large $y$, whose radial part has the form of approximately $\exp(\mi \kappa_0 y) / y$, with $\kappa_0^{2}$ the dimer binding energy ($\kappa_0>0$). Any incoming wave with a radial wave function of approximately $\exp(-\mi \kappa_0 y) / y$ must be canceled by adding appropriately the solution to the homogeneous part of \Eq{eqlambda}. This serves as the boundary condition at large $y$.

	\section{Perturbative method}
	
	When $\gamma$ is small, the solution can then be expressed as a series expansion $F(\lambda) = F^{0}(\lambda) + \gamma^{3/2} F^{1}(\lambda) +\cdots$, 
	where $F^{0}(\lambda)$ and $F^{1}(\lambda)$ satisfies
	\begin{align}
		\lambda^{-5} &= \Big( -1  + \hat{\mathbf{K}}_1 \Big) F^{0}(\lambda), \\
		0 &= \Big( -1  + \hat{\mathbf{K}}_1 \Big) F^{1}(\lambda) + \hat{\mathbf{A}}_0 F^{0}(\lambda). 
	\end{align}
	For a particular solution, we find the closed-form expressions (see Appendix~\ref{sect:pwave})
	\begin{align}
		F^{0}(\lambda)&=\frac{1}{8} \Big[ \mathrm{Ci}(\lambda) j_1(\lambda)+ \mathrm{Si}(\lambda) y_1(\lambda) - \frac{2}{\lambda^3} \Big] , \label{F0lambda} \\
		F^{1}(\lambda) 
		&=  -\frac{2}{15\sqrt{3}} \Big[ \mathrm{Si}(\lambda) j_1(\lambda)- \mathrm{Ci}(\lambda) y_1(\lambda) -\frac{1}{\lambda^2} - \frac{12}{ \lambda^{4}}  \Big] \nn\\
		& -\frac{\pi}{32}\cos\lambda + \mathcal{S}(\lambda), \label{F1lambda}
	\end{align}
	where $\mathrm{Ci}(\lambda) \equiv -\int_\lambda^{\infty} \dif t ~\cos(t)/t $ and $\mathrm{Si}(\lambda) \equiv \int_0^{\lambda} \dif t ~\sin(t)/t  $ are the cosine and sine integral functions, respectively, $\mathcal{S}(\lambda)$ is a smooth and non-oscillatory function, and $\mathcal{S}(\lambda) \sim \lambda^{-4} \ln \lambda$ for $\lambda\to +\infty $ and behaves as $\ln \lambda$ for $\lambda\to 0 $. 
	
	We further obtain the perturbative solution to the homogeneous part of \Eq{eqlambda}, which consists of a linear combination of two independent solutions $F_{\rm homo}(\lambda) = d_1 [j_1(\lambda) + \gamma^{3/2} (1-\hat{\mathbf{K}}_1)^{-1}\hat{\mathbf{A}}_0 j_1(\lambda) ] + d_2 [ y_1(\lambda) + \gamma^{3/2} (1-\hat{\mathbf{K}}_1)^{-1}\hat{\mathbf{A}}_0 y_1(\lambda) ]$ (see Appendix~\ref{sect:pwave}).
	We choose the coefficients $d_1$ and $d_2$ such that $f_\mu(\vect{y})$ satisfies the boundary conditions at small $y$ and at large $y$. 
	Then the final perturbative solution is expressed as $F(\lambda) = F_{\rm homo}(\lambda) + F^{0}(\lambda) + \gamma^{3/2} F^{1}(\lambda)$. 
	At large $\lambda$, $F(\lambda)$  should only contain the outgoing atom-dimer wave $ F(\lambda) \approx  c_{\rm ad} h_1^{(1)}(\lambda) $ with $h_l^{(1)}$ the spherical Hankel function of the first kind. 
	Thus, we find 
	\begin{equation}
		\label{cad}
		c_{\rm ad} \approx  \mi \frac{ \frac{1}{A_0(2)} + \frac{1}{4r_\ast} -\frac{8}{5\sqrt{3} r_\ast^{2}} }{1 + \left(-\frac{32}{3\sqrt{3}} + \frac{12}{\pi} \right) \frac{1}{r_\ast}  }  \gamma^{-3/2} + O\left(\gamma^{0} \right),
	\end{equation}
	where the dimensionless parameter $r_\ast \equiv y_c/R \gtrsim 1$.

	\section{Recombination rate constant}
	We calculate the recombination rate constant from the flux of the atom-dimer function. 
	At large $y$ and fixed $x$, we expect that the three-body wave function has the form
	\begin{equation}
		\label{pwavedimer}
		\psi \simeq \sum_{\mu=1}^{3} \psi_{\rm ad}^{(\mu)}(\vect{r}) \phi_{\rm 2b}^{(\mu)}(\vect{x}), 
	\end{equation}
	where $\vect{r}=\sqrt{3}\vect{y}/2$ is the atom-dimer relative position vector,
	$\psi_{\rm ad}^{(\mu)}(\vect{r})$ is the outgoing atom-dimer wave function, and 
	\begin{equation}
		\phi_{\rm 2b}^{(\mu)}(\vect{x}) = \frac{(1+\kappa_0 x) \me^{-\kappa_0 x} }{ (\frac{1}{R} -\frac{3}{2}\kappa_0)^{1/2} x^2 } Y^{(\mu)}(\hat{\vect{x}})
	\end{equation} 
	is the normalized wave function of the shallow dimer~\cite{Pricoupenko2006}.
	The wave number $\kappa_0$ satisfies ${1}/{v} -{\kappa_0^2}/{R}  +\kappa_0^3 =0$, and at large and positive $v$, $\kappa_0 = l_{\rm d}^{-1}[1 +\gamma^{3/2}/2 + 5\gamma^{3}/8 + O( \gamma^{9/2} )]$. 
	For the $p$-wave channel, the atom-dimer wave function has the form
	\begin{equation}
		\label{pwaveatomdimer}
		\psi_{\rm ad}^{(\mu)}(\vect{r}) = \sum_{\nu=1}^{3} \tilde{c}_{\mu\nu} h_1^{(1)}\left(\frac{2}{\sqrt{3}}\kappa_0r \right)Y^{(\nu)}(\hat{\vect{r}}).
	\end{equation}
	Comparing its large $y$ behavior to Eqs.~\eq{bc} and \eq{scheqcom} when $x\to 0$, we find 
	\begin{equation}
		f_{\mu}(\vect{y}) \approx {4 \pi}\left({\frac{1}{R} - \frac{3\kappa_0}{2}}\right)^{-1/2} \psi_{\rm ad}^{(\mu)}({\sqrt{3}}\vect{y}/2)
	\end{equation} 
	Then the coefficient $\tilde{c}_{\mu\nu}$ can be obtained from the large $y$ behavior of $f_{\mu\nu}(y)$ [see Eqs.~\eq{exp_fmu},  \eq{Fexpansion}, and \eq{auxf}].

\begin{figure}[ht!]
		\includegraphics[width=0.9\linewidth]{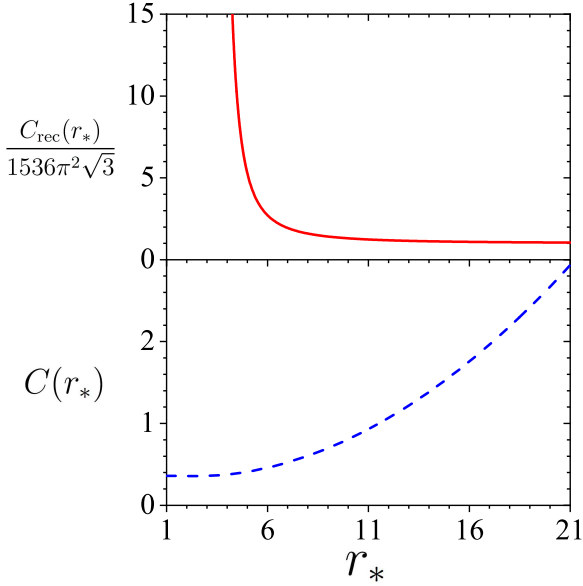}
		\caption{Plots of $C_{\rm rec}(r_\ast)$ (red solid curve) and $C(r_\ast)$ (blue dashed curve) as a function of $r_\ast$. Here $C_{\rm rec}(r_\ast)$ is normalized by $1536\pi^2\sqrt{3}$, the value of $C_{\rm rec}(r_\ast)$ when $r_\ast\to \infty$.}
		\label{fig:Crec}
	\end{figure}

	We define the recombination rate constant $\alpha_{\rm rec}$ such that the number of recombination events per unit time per unit volume is $(N/V)^{3}\alpha_{\rm rec}$, with $N$ the number of fermionic atoms. 
	We calculate the total flux of the atom-dimer outgoing wave, $\Phi = 3 V \lim_{r\to\infty} \int \dif \Omega_r ~ r^2 \hat{\vect{r}}\cdot \vect{j}(r)$.
	The factor $3$ appears because there are three output atom-dimer channels due to permutation and they contribute equally to the total flux. The $V$ comes from the integration over the center-of-mass position.
	The probability current $\vect{j}(r) = \frac{1}{\mi (2/3)} \mim \sum_{\mu=1}^{3}[ \psi_{\rm ad}^{(\mu)\ast}(\vect{r}) \nabla \psi_{\rm ad}^{(\mu)}(\vect{r})]$, where $2/3$ is the atom-dimer reduced mass. 
	From \Eq{pwaveatomdimer} we obtain 
	\begin{equation}
		\Phi =\frac{9\sqrt{3}V}{4  \kappa_0} \sum_{\mu,\nu=1}^{3} |\tilde{c}_{\mu\nu}|^2.
	\end{equation} 
	Also, according to the definition and a similar argument in Ref.~\cite{Mora2011}, $(N/V)^{3} \alpha_{\rm rec} = (\Phi/V) N(N-1)(N-2)/6$. The factor ${N(N-1)(N-2)}/{6}$ is the number of triplets among $N$ fermionic atoms.
	As $N$ is usually large, we take $N(N-1)(N-2) \approx N^3$ and therefore $\alpha_{\rm rec} = \frac{1}{6}V^2 \Phi$. 
	Finally, we obtain the leading-order result 
	\begin{equation}
		\alpha_{\rm rec} = C_{\rm rec}(r_\ast) \frac{\hbar}{M} v^{5/2} R^{1/2} |\vect{b}\times\vect{B}|^2. 
	\end{equation}
	Here and in the following similar formulas, the SI units are restored. The term
	\begin{equation}
		C_{\rm rec}^{}(r_\ast^{}) =  18\sqrt{3}\pi^2[A_0^{}(-1) A_0^{}(2)]^2\left(\frac{1}{\kappa_0^{} l_{\rm d}^{}} -\frac{3}{2} \gamma^{3/2} \right)\gamma^{3} |c_{\rm ad}^{}|^2 
	\end{equation}
	depends on the position of short-range boundary condition $r_\ast = y_c/R$ via $c_{\rm ad}$ given in \Eq{cad} and is plotted in \Fig{fig:Crec}.

	First, we note that $\alpha_{\rm rec} \propto v^{5/2} R^{1/2}$, consistent with the scaling predicted by the two-channel model calculation~\cite{Jona-Lasinio2008}. 
	At $R \ll y_c \ll l_{\rm d}$ ($1 \ll r_\ast \ll \gamma^{-3/2}$) we find $C_{\rm rec}(r_\ast) \approx 18\sqrt{3}\pi^2[A_0(-1)]^2+ O\big(r_\ast^{-2}\big)  \approx  1536\pi^2\sqrt{3}$, which remarkably recovers Eq.~(90) in Ref.~\cite{Jona-Lasinio2008}.

	We also see that $\alpha_{\rm rec} \propto |\vect{b}\times\vect{B}|^2$, which recovers the $E^2$ threshold law~\cite{Esry2001}. We can further obtain experimentally a more accessible recombination rate constant by averaging over the wave vectors $(\vect{b}, \vect{B})$ assuming a Maxwell Boltzmann distribution (see Appendix~\ref{sect:ta}) 
	\begin{equation}
		\langle\alpha_{\rm rec}\rangle_T  = \frac{3 \hbar}{8\pi^2 M} C_{\rm rec}(r_\ast) v^{5/2} R^{1/2} k_T^{4}, 
	\end{equation} 
	where $k_T =\sqrt{ 2\pi M k_B T  /\hbar^2}$ (defined as $2\pi$ over the thermal de Broglie wavelength), $k_B$ is the Boltzmann constant, and $T$ is the temperature.

	\section{Subleading correction}
	
	Now let us consider the subleading correction to $\alpha_{\rm rec}$. 
	Equation \eq{eqf} can be expanded perturbatively at small $E$. In the $p$-wave channel of the atom-dimer motion, the subleading order gives an $E^3$ correction to $\alpha_{\rm rec}$. Meanwhile, according to the threshold law~\cite{Esry2001}, the $s$- and $d$-wave channels also give the same $E^3$ contributions. Thus, the subleading term consists of these three parts.

	\subsection{Subleading order in the $p$-wave channel}
	\label{sect:subpwave}
	
	First, we consider the contribution from the $p$-wave channel. We expand \Eq{eqf} at small $\kappa^2$ and consider the equation for $h_{\mu\nu}(y)$ defined in \Eq{Fexpansion}. 
	Because the atom-dimer flux is proportional to $\sum_{\mu,\nu=1}^{3} |\tilde{c}_{\mu\nu}|^2 $ for the leading-order approximation and $f_{\mu\nu}(y)$ is antisymmetric in the two indices $\mu$ and $\nu$, only the anti-symmetric part $h_{\mu\nu}^{A}(y) \equiv \frac{1}{2} [ h_{\mu\nu}(y) -h_{\nu\mu}(y) ]$ contributes to the subleading correction.
	We obtain
	\begin{equation}
		\label{eqsubpwave}
		\frac{12\pi}{\kappa^2} D_{\mu\nu}^{1\rm A}  y^3  =\Big( -\frac{1}{v} + \frac{\hat{K}_1}{R}  + \hat{A}_0 \Big) h_{\mu\nu}^{\rm A}(y)  +  \Big(\frac{1}{R} + \hat{A}_1 \Big)f_{\mu\nu}(y), 
	\end{equation}
	where $D_{\mu\nu}^{1\rm A} \equiv (D_{\mu \nu}^{1} - D_{\nu \mu}^{1}) / 2 = \kappa^2 D_{\mu \nu}^{0} / 20$.
	Here $1\le \mu, \nu \le 3$ label the $p$-wave components. In addition, $\hat{A}_1$ is an integral operator (see Appendix~\ref{sect:operators}). 
	At $-2<\mre (s)<4$, $\hat{A}_1 y^{-s} = A_1(s) y^{-s-1} $, where
	\begin{align}
		A_1(s) &= \frac{1}{(s-1)\sin\frac{\pi s}{2}} \Big[ 4\sqrt{3}(s-4) \sin \frac{\pi s}{6} + 12 s \cos \frac{\pi s}{6} \nn\\
		&\qquad\qquad\qquad\qquad - \frac{3}{2} s (s-2) \cos \frac{\pi s}{2} \Big]. 
	\end{align}

	We define an auxiliary function $H(\lambda)$ such that 
	\begin{align}
		h_{\mu\nu}^{\rm A}(y) = \frac{12\pi}{\kappa^2} D_{\mu\nu}^{1\rm A} v l_{\rm d}^3 \Big[ -\lambda^3 + 12\lambda + w_1 \gamma^{3/2} \nn\\
		+ w_2 \gamma^{3/2} \lambda^{-2} +  w_2 \gamma^{3}A_0(2) H(\lambda)  \Big] 
	\end{align}
	and obtain the dimensionless form of the equation
	\begin{align}
		\label{eqH}
		\Big(-1+\hat{\mathbf{K}}_1 +  \gamma^{3/2} \hat{\mathbf{A}}_0 \Big) H(\lambda)  = w_3\lambda^{-3} - \lambda^{-5} \nn\\
		- \frac{20A_0(-1)}{w_2} \Big(1+ \gamma^{3/2} \hat{\mathbf{A}}_1\Big) F(\lambda), 
	\end{align}
	where $w_1 \equiv 2A_1(-1) -A_0(-3) $,  $w_2 \equiv 4 A_1(-1)  - 2A_0(-3) +14 A_0(-1) $, and $w_3 \equiv - [w_1 A_0(0) + 2A_0(-1) A_1(2)] / [w_2 A_0(2)] $.

	We solve \Eq{eqsubpwave} perturbatively at small $\gamma$. Similarly to the leading contribution, to uniquely fix the solution, we impose the following boundary conditions: (i) At short distances $y\le y_c = r_\ast R$, $h_{\mu\nu}^{\rm A}(y) = 0$ and (ii) at large distances, $H(\lambda) \approx  c_{\rm ad}^{\rm 1A}  h_1^{(1)}(\lambda)$. 
	Here the coefficient $c_{\rm ad}^{\rm 1A}$ describes the amplitude of the atom-dimer outgoing wave.
	Note that in \Eq{eqH}, the first source term $w_3\lambda^{-3}$ and third source terms $-[20 A_0(-1) / w_2] (1+ \gamma^{3/2} \hat{A}_1 ) F(\lambda)$ are much less important than the second term $-\lambda^{-5}$. 
	The small-distance behavior of a particular solution of $H(\lambda)$ is mainly determined by the second source term $-\lambda^{-5}$. 
	Then, according to \Eq{eqlambda}, we obtain $c_{\rm ad}^{\rm 1A} \approx  - c_{\rm ad}$. 
	As a result, we find the correction
	\begin{equation}
		\Delta\alpha_{\rm rec}^{l=1}  \approx  - \frac{ 9 }{ 20 } \kappa^2 l_{\rm d}^{2} \alpha_{\rm rec},
	\end{equation}
	and after a thermal averaging, 
	\begin{equation}
		\langle\Delta\alpha_{\rm rec}^{l=1}\rangle_T  \approx  \frac{27 \hbar}{64\pi^3 M} C_{\rm rec}(r_\ast) v^{5/2} R^{1/2} l_{\rm d}^{2} k_T^{6},
	\end{equation}
	where the wave number $k_T =\sqrt{ 2\pi M k_B T  /\hbar^2}$ (defined as $2\pi$ over the thermal de Broglie wavelength), $k_B$ is the Boltzmann constant, and $T$ is the temperature.

	\subsection{Contribution from the $s$-wave channel}
	\label{sect:swave}
	Now let us consider the contributions from the $s$- and $d$-wave channels. To obtain analytic results, we will neglect the coupling between the $s$- and $d$-wave channels in \Eq{eqf}.
	First we consider the $s$-wave channel and assume $f_\mu(\vect{y}) = f_\mu(y)$. 
	At low energies, from \Eq{eqf}, we find
	\begin{equation}
		12\pi \overbar{C}_{\mu}^{l=0} y^2 = \Big( -\frac{1}{v} + \frac{1}{R} \hat{K}_0 + \hat{L}_0^{l=0} +\hat{S}_0^{l=0} \Big) f_{\mu}(y),
	\end{equation}
	where $\hat{L}_0^{l=0}$ and $\hat{S}_0^{l=0}$ are the $s$-wave component of the operators defined in Eqs.~\eq{Loperator} and \eq{Soperator} (see Appendix~\ref{sect:operators}) and $\overbar{C}_{\mu}^{l=0} = -\frac{\sqrt{2\pi} }{9V^{3/2}}\mi \sum_{i=1}^{3} b_i^{(\mu)} B_i^2$.

	We define the auxiliary function $F^{l=0}(\lambda)$ such that
	\begin{align}
		f_{\mu}(y) &= 12\pi \overbar{C}_{\mu}^{l=0} v l_{\rm d}^{2} \Big[-\lambda^2 + 6 -\gamma^{3/2} A_{0}^{l=0}(-2) \lambda^{-1} \nn\\
		&\qquad\qquad + \gamma^3 A_{0}^{l=0}(1) A_{0}^{l=0}(-2)  F^{l=0}(\lambda) \Big], 
	\end{align}
	and $F^{l=0}(\lambda)$ satisfies
	\begin{equation}
		\label{eqFswave}
		\lambda^{-4} = \Big(-1 + \hat{\mathbf{K}}_{0} + \gamma^{3/2} \hat{\mathbf{A}}_0^{l=0} \Big) F^{l=0}(\lambda),
	\end{equation}
	where $\hat{\mathbf{A}}_0^{l=0} \equiv l_{\rm d}^3 ( \hat{L}_0^{l=0} +\hat{S}_0^{l=0} )$.
	At $-5< \mre(s) <5$, $\hat{\mathbf{A}}_0^{l=0} \lambda^{-s} = A_0^{l=0}(s) \lambda^{-s-3}$, where
	\begin{align}
		A_0^{l=0}(s) & =\frac{1}{\cos \frac{\pi s}{2}} \Big[-s(s^2-1) \sin \frac{\pi s}{2}  - 8s \sin \frac{\pi s}{6} \nn\\
		& + \frac{8}{\sqrt{3}} s^2 \cos \frac{\pi s}{6} \Big]. 
	\end{align}

	We solve \Eq{eqFswave} perturbatively at small $\gamma$ in a similar way.
	To uniquely fix the solution, we impose the following boundary conditions: (i) At short distances $y\le y_c = r_\ast R$, $f_{\mu}(y) = 0$, and (ii) at large distances, $F^{l=0}(\lambda) \approx  c_{\rm ad}^{l=0}  h_1^{(1)}(\lambda)$. 
	Here the coefficient $c_{\rm ad}^{l=0} $ describes the amplitude of the atom-dimer outgoing wave. 
	Then we find 
	\begin{align}
		c_{\rm ad}^{l=0} &= -\mi \frac{6r_\ast }{A_0^{l=0}(1) A_0^{l=0}(-2)} \Bigg[ 1-  \frac{A_0^{l=0}(-2)}{6 r_\ast} \nn\\
		&- \frac{A_0^{l=0}(1) A_0^{l=0}(-2)}{12 r_\ast^{2}} -  \frac{  A_0^{l=0}(1) A_0^{l=0}(-2)}{ 9 \sqrt{3} r_\ast^{3}} \Bigg] \nn\\
		& \times \Bigg[1+ \left(\frac{8}{3\sqrt{3}} -\frac{2}{\pi}\right) \frac{1}{r_\ast}   \Bigg]^{-1} \gamma^{-3/2} +O(\gamma^0 ).
	\end{align}
	Also, we know that in the $s$-wave channel the atom-dimer outgoing wave $\psi_{\rm ad}^{(\mu)}(\vect{r}) =  \tilde{c}_{\mu} h_{0}^{(1)} ( 2\kappa_0 r /\sqrt{3} )$, 
	with $\tilde{c}_{\mu} = \frac{ \sqrt{\frac{1}{R} -\frac{3}{2}\kappa_0} }{4\pi }  12\pi \overbar{C}_{\mu}^{l=0} v l_{\rm d}^{2}  \gamma^3 A_{0}^{l=0}(1) A_{0}^{l=0}(-2)    c_{\rm ad}^{l=0}$, 
	and then the atom-dimer flux $\Phi = [ 9\sqrt{3} \pi \hbar V / (M \kappa_0) ] \sum_{ \mu =1}^{3} |\tilde{c}_{\mu}|^2 $. 
	Finally, we obtain its contribution to the recombination rate constant
	\begin{align}
		\Delta \alpha_{\rm rec}^{l=0} &\approx C_{\rm rec}^{l=0}(r_\ast) \frac{\hbar}{M} v^{5/2}R^{1/2} l_{\rm d}^{2} \Big[b^2 (b^2-B^2)^2 \nn\\
		&\qquad\qquad\qquad\qquad + 4 b^2 B^2 (2B^2 - b^2) t^2\Big],
	\end{align} 
	where $t\equiv \hat{\vect{b}}\cdot \hat{\vect{B}}$ and
	\begin{align}
		C_{\rm rec}^{l=0}(r_\ast) & \approx  \frac{27\sqrt{3}\pi^2}{4}   r_\ast^{2} \Big[1-  \frac{A_0^{l=0}(-2)}{6 r_\ast} \nn\\ 
		&- \frac{A_0^{l=0}(1) A_0^{l=0}(-2)}{12 r_\ast^{2}} -  \frac{  A_0^{l=0}(1) A_0^{l=0}(-2)}{ 9 \sqrt{3} r_\ast^{3}} \Big]^2 \nn\\
		&\left[ 1+ \left(\frac{8}{3\sqrt{3}} -\frac{2}{\pi} \right) \frac{1}{r_\ast} \right]^{-2}. 
	\end{align}
	After thermal averaging, 
	\begin{equation}
		\langle \Delta\alpha_{\rm rec}^{l=0} \rangle_T \approx \frac{15 \hbar}{8\pi^3 M} C_{\rm rec}^{l=0}(r_\ast)  v^{5/2}R^{1/2} l_{\rm d}^{2} k_T^{6}. 
	\end{equation}

	\subsection{Contribution from the $d$-wave channel}
	\label{sect:dwave}
	Finally, let us consider the $d$-wave channel and assume $f_\mu(\vect{y}) = \sum_{i=1}^{5} f_{\mu i}(y) D_i(\hat{\vect{y}})$, where $D_i(\vect{y})$ is the real $d$-wave spherical harmonics defined as
	\begin{align}
		\mathcal{D}_1(\hat{\vect{y}}) &= \sqrt{\frac{15}{4\pi}} \frac{y^{(1)} y^{(2)}}{y^2}, \nn\\
		\mathcal{D}_2(\hat{\vect{y}}) &= \sqrt{\frac{15}{4\pi}} \frac{y^{(2)} y^{(3)}}{y^2}, \nn\\
		\mathcal{D}_3(\hat{\vect{y}}) &= \sqrt{\frac{15}{4\pi}} \frac{y^{(3)} y^{(1)}}{y^2}, \nn\\
		\mathcal{D}_4(\hat{\vect{y}}) &= \sqrt{\frac{5}{16\pi}} \Big[3 \Big( \frac{  y^{(3)} }{y} \Big)^2 -1 \Big] ,\nn\\
		\mathcal{D}_5(\hat{\vect{y}}) &= \sqrt{\frac{15}{16\pi}} \Big[ \Big( \frac{  y^{(1)} }{y} \Big)^2 -\Big( \frac{  y^{(2)} }{y} \Big)^2 \Big] , \nn
	\end{align}
	where $\vect{y}\equiv ( y^{(1)}, y^{(2)}, y^{(3)}  )$, $y \equiv |\vect{y}|$ and $\hat{\vect{y}} \equiv \vect{y} /y$. 
	Here $1\le \mu \le 3$ and $1\le i \le 5$ label the $p$- and $d$-wave components, respectively. 
	At low energies, neglecting the mixing between the $s$-wave and the $d$-wave channels, from \Eq{eqf} we find, 
	\begin{align}
		12\pi \overbar{C}_{\mu i}^{l=2} y^2 &= \Big( -\frac{1}{v} +\frac{1}{R} \hat{K}_2 + \hat{L}_0^{l=2} \Big) f_{\mu i}(y) \nn\\
		& + \sum_{\nu=1}^{3}\sum_{j=1}^{5} \hat{S}_{\mu i, \nu j} f_{\nu j}(y), 
	\end{align}
	where $\overbar{C}_{\mu i}^{l=2} = -\mi \frac{8\sqrt{2} \pi^{3/2} }{45 V^{3/2}} \sum_{k=1}^{3} b_k^{(\mu)} B_k^{2} \mathcal{D}_i(\hat{\vect{B}}_k)$. 
	The operator $\hat{S}_{\mu i, \nu j}$ can be expressed in the form 
	\begin{equation}
		\hat{S}_{\mu i, \nu j} = \widetilde{A}_{\mu i, \nu j} \hat{S}^{\rm A} + \widetilde{B}_{\mu i, \nu j} \hat{S}^{\rm B}, 
	\end{equation} 
	where $\widetilde{A}_{\mu i, \nu j}$ and $ \widetilde{B}_{\mu i, \nu j} $ are numerical coefficients and $\hat{S}^{\rm A}$ and $\hat{S}^{\rm B}$ are integral operators defined as
	\begin{align}
		\hat{S}^{\rm A} f(y) &= \int_{0}^{\infty}\dif y'\int_{-1}^{1}\dif \tau~ f(y') \frac{288 y y'^3}{\chi^8} P_1(\tau), \\
		\hat{S}^{\rm B} f(y) &= \int_{0}^{\infty}\dif y'\int_{-1}^{1}\dif \tau~ f(y') \frac{288 y y'^3}{\chi^8} P_3(\tau),
	\end{align}
	where $P_l(\tau)$ is the Legendre polynomial and $\tau \equiv \cos \Theta = \frac{\vect{y}\cdot\vect{y}'}{y y'} $ with $\Theta$ the angle between $\vect{y}$ and $\vect{y}'$. 
	To diagonalize the equation, we use the collective indices $\bar{i}\equiv \mu i $ and $\bar{j}\equiv \nu j$ (in a proper order $1\le \bar{i}, \bar{j}\le 15$). 
	Then we make the transforms 
	\begin{align} 
		\hat{S}_{\bar{i}\bar{j}} &= \Big( Q \hat{S}^{\rm D} Q^{-1}\Big)_{\bar{i}\bar{j}}, \\
		\widetilde{f}_{\bar{i}}(y) &= \sum_{\bar{j}} \Big( Q^{-1} \Big)_{\bar{i}\bar{j}} f_{\bar{j}}(y),\\ 
		\widetilde{C}_{\bar{i}}^{l=2} &= \sum_{\bar{j}} \Big( Q^{-1} \Big)_{\bar{i}\bar{j}} \overbar{C}_{\bar{i}}^{l=2}, 
	\end{align} 
	where $Q$ is a matrix of real numbers, $\hat{S}^{\rm D} $ is a diagonal matrix 
	\begin{align}
		\hat{S}_{\bar{i} \bar{i}}^{\rm D} = 
		\begin{cases}
			-\frac{3}{20\pi}( \hat{S}^{\rm A}- \hat{S}^{\rm B} ),  & \qquad 1\le \bar{i} \le 5 \\
			\frac{3}{140\pi}( 14 \hat{S}^{\rm A} + \hat{S}^{\rm B} ),  & \qquad 6\le \bar{i} \le 12 \\
			\frac{1}{20\pi}( \hat{S}^{\rm A} + 9 \hat{S}^{\rm B} ),  & \qquad 13\le \bar{i} \le 15, \\
		\end{cases} 
	\end{align} 
	and $\widetilde{C}_{\bar{i}}^{l=2} = 0 $ when $ 1\le \bar{i} \le 5$. 
	Then we obtain the diagonalized equation
	\begin{equation}\label{dwave}
		12\pi \widetilde{C}_{\bar{i}}^{l=2} y^2  = \Big( -\frac{1}{v} +\frac{1}{R}  \hat{K}_2 + \hat{L}_0^{l=2} + \hat{S}_{\bar{i}\bar{i}}^{\rm D} \Big) \widetilde{f}_{\bar{i}}(y).
	\end{equation}
	Similarly, we define the auxiliary function $F_{\bar{i}}^{l=2}(\lambda)$ such that
	\begin{align}
		\widetilde{f}_{\bar{i}}(y) & = 12\pi \widetilde{C}_{\bar{i}}^{l=2} v l_{\rm d}^{2} \Big[-\lambda^2  -\gamma^{3/2} W_{\bar{i}}(-2) \lambda^{-1} \nn\\
		&-6 \gamma^{3/2} W_{\bar{i}}(-2) \lambda^{-3} + 6 \gamma^3 W_{\bar{i}}(3) W_{\bar{i}}(-2) F_{\bar{i}}^{l=2}(\lambda) \Big], 
	\end{align}
	and $F_{\bar{i}}^{l=2}(\lambda)$ satisfies
	\begin{equation}
		\label{eqFdwave}
		\lambda^{-6} + \frac{W_{\bar{i}}(1)}{6 W_{\bar{i}}(3)} \lambda^{-4} = \Big(-1 + \hat{\mathbf{K}}_{2} + \gamma^{3/2} \hat{W}_{\bar{i}} \Big) F_{\bar{i}}^{l=2}(\lambda),
	\end{equation}
	where $\hat{W}_{\bar{i}} \equiv l_{\rm d}^{3} \Big( \hat{L}_0^{l=2} + \hat{S}_{\bar{i}\bar{i}}^{\rm D} \Big) $, and $\hat{W}_{\bar{i}} \lambda^{-s} =	W_{\bar{i}}(s) \lambda^{-s-3}$ at $-5\le \mre (s) \le 5$.

	We solve \Eq{eqFdwave} perturbatively at small $\gamma$ in a similar way.
	To uniquely fix the solution, we impose the following boundary conditions: (i) At short distances $y\le y_c = r_\ast R$, $f_{\mu}(\vect{y}) = 0$, and (ii) at large distances, $F_{\bar{i}}^{l=2}(\lambda) \approx  c_{{\rm ad}, \bar{i}}^{l=2}  h_2^{(1)}(\lambda)$. 
	Here the coefficient $c_{{\rm ad}, \bar{i}}^{l=2} $ describes the amplitude of the atom-dimer outgoing wave. 
	For simplicity, we only consider the $\gamma^{0}$th-order solution to \Eq{eqFdwave}. 
	Then we find 
	\begin{equation}
		c_{{\rm ad}, \bar{i}}^{l=2} \approx  \mi \gamma^{-3/2} \Big[ \frac{1}{3 W_{\bar{i}}(3) }  + \frac{1}{18 r_\ast} \Big].
	\end{equation}
	In the $d$-wave channel, the atom-dimer outgoing wave $\psi_{\rm ad}^{(\mu)}(\vect{r}) =  \sum_{i=1}^{5} \tilde{c}_{\mu i} h_{2}^{(1)}( 2 \kappa_0 r / \sqrt{3} ) \mathcal{D}_i(\hat{\vect{r}})$ with $\tilde{c}_{\mu i} = \frac{ \sqrt{\frac{1}{R} -\frac{3}{2}\kappa_0} }{4\pi } \sum_{\bar{j}} P_{\bar{i}\bar{j}} [ 12\pi \widetilde{C}_{\bar{j}}^{l=2} v l_{\rm d}^{2} 6 \gamma^3 W_{\bar{j}}(3) W_{\bar{j}}(-2) c_{{\rm ad}, \bar{j}}^{l=2}]$ and the collective index $\bar{i} = \mu i $. Then the atom-dimer flux $\Phi = [9\sqrt{3}  \hbar V / (4 M \kappa_0)] \sum_{ \mu =1}^{3} \sum_{ i =1}^{5} |\tilde{c}_{\mu i}|^2 $. 
	Finally, we find the $d$-wave contribution to the recombination rate constant
	\begin{align}
		\Delta\alpha_{\rm rec}^{l=2} &\approx \frac{\hbar}{M} v^{5/2}R^{1/2} l_{\rm d}^{2}\Biggl\{  \frac{992\sqrt{3} \pi^2}{49}\Big[b^6 \nn\\
		&\mkern-20mu +\frac{2}{115} b^4 B^2 (208-673 t^2) + \frac{1}{155} b^2 B^4 (881 + 514 t^2)\Big] \nn\\
		&\mkern-20mu +  \frac{\#_1}{r_\ast} + \frac{\#_2}{r_\ast^2} \Biggr\},
	\end{align} 
	where $t\equiv \hat{\vect{b}}\cdot \hat{\vect{B}}$ and $\#_{1,2}$ represents two functions of $(b,B,t)$. 
	After a thermal averaging,
	\begin{equation}
		\langle\Delta\alpha_{\rm rec}^{l=2}\rangle_T \approx  \frac{\hbar}{8\pi^3 M} C_{\rm rec}^{l=2}(r_\ast) v^{5/2}R^{1/2} l_{\rm d}^{2} k_T^{6},
	\end{equation} 
	where 
	\begin{align} 
		& C_{\rm rec}^{l=2}(r_\ast) \!  \approx \! \frac{7104\sqrt{3}}{7}\pi^2 + \! \Big( \!\! - \frac{313344\sqrt{3} \pi}{ 49} + \frac{309248\pi^2}{147}  \Big) \frac{1}{r_\ast} \nn\\
		& \qquad + \Big( \frac{16957440\sqrt{3}}{343} - \frac{18776064\pi}{343}   + \frac{146243584 \pi^2}{9261 \sqrt{3}}\Big) \frac{1}{r_\ast^2}. 
	\end{align} 
	
	\subsection{Overall contributions to three-body relaxation rate}
	\label{sect:totalresult}
	Now we can collect results from all partial waves, including both the leading and subleading contributions, and obtain the final result for the recombination rate constant in the form
	\begin{equation}
		\langle\alpha_{\rm rec}\rangle_T  = \frac{3\hbar}{8\pi^2M} C_{\rm rec}(r_\ast) v^{5/2} R^{1/2} k_T^{4}\Big[1+C(r_\ast)k_T^2l_{\rm d}^2\Big],
	\end{equation} 
	where the coefficient $C(r_\ast)$ is given by
	\begin{equation}
		C(r_\ast)=C^{-1}_{\rm rec}(r_\ast)\left[\frac{9}{8\pi}C_{\rm rec}(r_\ast) + \frac{5}{\pi}C_{\rm rec}^{l=0}(r_\ast) + \frac{1}{3\pi}C_{\rm rec}^{l=2}(r_\ast)\right].  
	\end{equation} 
	A numerical figure of $C(r_\ast)$ is plotted in \Fig{fig:Crec}. For reasonable choices of $r_*$, we note that $C(10) \approx 0.81$ and $C(15) \approx 1.56$; both are of order one.

	\section{Summary}
	
	We studied the recombination of three identical fermionic atoms with $p$-wave interaction into a shallow $p$-wave dimer state. Our results highlight the importance of the $p$-wave effective range in three-body observables with the leading-order term scaling as $v^{5/2}R^{1/2}$. The subleading terms from $s$-, $p$-, and $d$-wave channels are also calculated and their corrections are characterized by the dimensionless parameter $k_T^2l_{\rm d}^2$. Given a reasonable value of $r_*\approx 15$, then we expect that the subleading term will become important when $k_T^2l_{\rm d}^2\gtrsim 0.64$, which is experimentally quite accessible. In fact, the experiment~\cite{Yoshida2018} conducted on the $v<0$ side indicated that such deviation occurred when $k_T^2l_{\rm d}^2\sim 0.4$, which is equivalent to the parameter $k_ek_T^2V_B\sim 0.095$ used in Ref.~\cite{Yoshida2018}.

	\begin{acknowledgments}
		S. Zhu and S. Zhang were supported by HK GRF Grants No. 17318316 and No. 17305218, CRF Grants No. C6026-16W and N0. C6005-17G, and the Croucher Foundation under the Croucher Innovation Award. Z.Y. was supported by the Key Area Research and Development Program of Guangdong Province (Grant No. 2019B030330001), the National Natural Science Foundation of China (Grants No. 11474179, No. 11722438, No. 91736103, and No. 12074440), and the Guangdong Project (Grant No. 2017GC010613). 
	\end{acknowledgments}

	\appendix

	\section{Properties of the integral operators}
	\label{sect:operators}
	
	First we consider the integral operator $\hat{L}$, defined as
	\begin{align}
		\hat{L} f(\vect{y}) &= \frac{3 \kappa^6}{2\pi^2} \mathcal{P} \int \dif^3 y'~ \Big\{  \Big[f(\vect{y})-f(\vect{y}') \Big] g_3 \Big(\kappa \xi \Big) \nn\\
		& + g_3\Big(\kappa \chi\Big) f(\vect{y}')  \Big\}. 
	\end{align}
	We rewrite it in a partial wave expansion. 
	Let $f(\vect{y}) = \sum_{lm}  f_{lm}(y) Y_{lm}(\hat{\vect{y}})$. 
	Then, we find
	\begin{align}
		\hat{L} f(\vect{y})&= \sum_{lm} \Big[\hat{L}^{l}f_{lm}(y) \Big] Y_{lm}(\hat{\vect{y}}) \\ 
		\hat{L}^{l}f_{lm}(y)&=\frac{3 \kappa^6}{\pi} \!\! \int_{-1}^{1}\dif \tau~ \mathcal{P} \!\! \int_{0}^{\infty} \!\! \dif y' ~ y'^2  \Big\{   f_{lm}(y') P_{l}(\tau) g_3\Big(\kappa \chi \Big) \nn\\
		& + \Big[f_{lm}(y) - f_{lm}(y') P_{l}(\tau)   \Big]g_3 \Big(\kappa \xi \Big) \Big\}, 
	\end{align} 
	where $P_l(\tau)$ is the Legendre polynomial and $\tau \equiv\cos \Theta = \frac{\vect{y}\cdot\vect{y}'}{y y'} $, with $\Theta$ the angle between $\vect{y}$ and $\vect{y}'$.
	Here and in the following we use the shorthand notation
	\begin{align}
		\xi &\equiv |\vect{y}' - \vect{y}|,  \qquad \chi \equiv \sqrt{y^2+y'^2+\vect{y}\cdot\vect{y}'}, \\
		\xi_\pm &\equiv |y\pm y'|, \qquad \chi_\pm \equiv \sqrt{y^2+y'^2 \pm y y'}. 
	\end{align} 
	We see that $\hat{L}$ conserves the angular momentum $l$. 
	After integrating out the angular part we have the following expressions: For the $s$ wave,
	\begin{align}
		\hat{L}^{l=0} f(y) &= - \frac{3}{\pi } \kappa^4 \mathcal{P} \int_0^\infty \dif y'~ \frac{y'}{y} \sum_{\sigma=\pm} \sigma \Big\{ \nn\\
		&\Big[f(y)-f(y')\Big]  g_2(\kappa \xi_\sigma) + 2 f(y')  g_2(\kappa \chi_\sigma)  \Big\}, 
	\end{align}
	and for the $p$ wave, 
	\begin{align}
		\hat{L}^{l=1} f(y) &= - \frac{3}{\pi } \kappa^4 \mathcal{P}\int_{0}^{\infty}\dif y' ~ \frac{y'}{y}  \sum_{\sigma=\pm} \Big\{  \sigma  f(y)  g_2(\kappa \xi_{\sigma}) \nn\\
		& + f(y') \Big[ g_2(\kappa \xi_{\sigma}) +2 g_2(\kappa \chi_{\sigma}) + \frac{\sigma}{\kappa^2 y y'} g_1(\kappa \xi_{\sigma}) \nn\\
		&+  \frac{4\sigma}{\kappa^2 y y'} g_1(\kappa \chi_{\sigma})\Big] \Big\}.
	\end{align}
	We define the low-energy expansion
	\begin{equation}
		\hat{L}^{l} f(y) = \hat{L}_0^{l} f(y) + \kappa^2 \hat{L}_1^{l} f(y) +O(\kappa^4). 
	\end{equation}

	Now let us consider the operator $\hat{S}_{\mu}$, defined as
	\begin{equation}
		\hat{S}_{\mu} \vect{f}(\vect{y}) =  \frac{3\sqrt{3\pi} \kappa^8}{2\pi^2}  \int \dif^3 y'~  g_4\Big(\kappa \chi \Big) \Big[ \vect{y} \cdot \vect{f}(\vect{y}') \Big] y' Y^{(\mu)}(\hat{\vect{y}}').
	\end{equation}
	As the integral kernel contains a $p$-wave angular part $Y^{(\mu)}(\hat{\vect{y}}')$,  $\hat{S}_{\mu}$ does not conserve the angular momentum. 
	We see that $\hat{S}_{\mu}$ effectively adds an angular momentum $l=1$ to that of the atom-dimer function $\vect{f}(\vect{y})$. 
	For simplicity, we can find the proximate result of $\hat{S}_{\mu}$ by assuming $\vect{f}(\vect{y})$ to be in a specific partial wave channel and neglecting the mixing between different channels.

	Consider the $s$-wave channel and assume $\vect{f}(\vect{y}) =\vect{f}(y)$. 
	We find 
	\begin{equation}
		\hat{S}_{\mu} \vect{f}(y) = \hat{S}^{l=0} f_{\mu}(y) + \Big( \text{$d$-wave part} \Big), 
	\end{equation}
	where the operator $\hat{S}^{l=0}$ is defined as
	\begin{align}
		\hat{S}^{l=0} f(y) &= -\frac{3\kappa^6}{\pi} \int_0^\infty \dif y'~   y'^2 f(y') \sum_{\sigma=\pm} \Big[  g_3(\kappa \chi_{\sigma})  \nn\\
		&+ \frac{2\sigma }{\kappa^2 y y'}  g_2(\kappa \chi_{\sigma}) \Big].
	\end{align}
	We neglect the $d$-wave part and obtain $\hat{S}_{\mu} \vect{f}(y) \approx  \hat{S}^{l=0} f_{\mu}(y)$.
	We also define the low-energy expansion
	\begin{equation}
		\hat{S}^{l=0} f(y) = \hat{S}_{0}^{l=0} f(y) + \kappa^2 \hat{S}_{1}^{l=0} f(y) +O(\kappa^4). 
	\end{equation}

	Consider the $p$-wave channel and assume $f_\mu(\vect{y}) = \sum_{\nu=1}^{3} f_{\mu\nu}(y) Y^{(\nu)}(\hat{\vect{y}})$ with $\mu=1,2,3$. 
	We find
	\begin{align}
		\hat{S}_{\mu} \vect{f}(y) &= \sum_{\nu,\rho=1}^{3} \frac{3\kappa^8 y }{\pi}  Y^{(\nu)}(\hat{\vect{y}}) \int_{0}^{\infty}\dif y' \int \dif\Omega_{y'}~  \nn\\
		& \Big[ y'^3 f_{\nu\rho}(y') g_4(\kappa \chi )  Y^{(\rho)}(\hat{\vect{y}}')Y^{(\mu)}(\hat{\vect{y}}') \Big] \nn\\
		&=\sum_{\nu =1}^{3} Y^{(\nu)}(\hat{\vect{y}})  \Big\{3\hat{R}\Big[f_{11}(y)+f_{22}(y)+f_{33}(y)\Big]\delta_{\mu\nu} \nn\\
		&+3\hat{R} f_{\mu\nu}(y) + (\hat{T}-2\hat{R}) f_{\nu\mu}(y) \Big\}  \nn\\ 
		&+ \Big( \text{$f$-wave part} \Big),
	\end{align}
	where the integral operators $\hat{T}$ and $\hat{R}$ are defined as
	\begin{align}
		\hat{T} f(y) &= -\frac{3\kappa^6}{\pi} \int_{0}^{\infty} \dif y'~ y'^2 f(y') \sum_{\sigma=\pm} \sigma g_3(\kappa \chi_{\sigma}), \\
		\hat{R} f(y) &= -\frac{3\kappa^6 }{5\pi} \int_{0}^{\infty} \dif y'~  y'^2 f(y') \sum_{\sigma=\pm} \Big[ \sigma g_3(\kappa \chi_{\sigma}) \nn\\
		&+ \frac{6 }{\kappa^2 y y'} g_2(\kappa \chi_{\sigma}) + \frac{12 \sigma}{\kappa^4 y^2 y'^2}  g_1(\kappa \chi_{\sigma}) \Big]. 
	\end{align}
	We neglect the $f$-wave part and only retain the $p$-wave part of $\hat{S}_{\mu} \vect{f}(y)$. 
	Also, we define the low-energy expansion of the operators
	\begin{align}
		\hat{T} f(y) &= \hat{T}_0 f(y) + \kappa^2 \hat{T}_1 f(y) +O(\kappa^4), \\
		\hat{R} f(y) &= \hat{R}_0 f(y) + \kappa^2 \hat{R}_1 f(y) +O(\kappa^4).
	\end{align}
	Then we find that $\hat{A}_0 \equiv \hat{L}_{0}^{l=1} + 5\hat{R}_{0} - \hat{T}_{0}$ in \Eq{eqpwave} and $\hat{A}_1 \equiv \hat{L}_1^{l=1} +\hat{T}_1+\hat{R}_1$ in \Eq{eqsubpwave}.

	\section{Perturbative method in the $p$-wave channel}  
	\label{sect:pwave}

	First we seek a particular solution of \Eq{eqlambda}. 
	For the leading order term, we straightforwardly obtain the particular solution \Eq{F0lambda}.

	Consider the first-order term $F^{1}(\lambda) = ( 1 - \hat{\mathbf{K}}_1 )^{-1} \hat{\mathbf{A}}_0 F^{0}(\lambda)$. It is convenient to write the expressions in the form of the inverse Mellin transforms. 
	We have the following observations about the inverse Mellin transforms of $F^{0}(\lambda)$ and its momentum space counterpart $F_q^{0}$:
	\begin{equation}
		\widetilde{F}^{0}(s) = \int_0^{\infty}\dif q~ F_q^{0} q^{-s+2} = \frac{\pi}{16} \tan \frac{\pi s}{2},
	\end{equation}
	with $3<\mre (s)<5$; 
	\begin{equation}
		F_q^{0} = \int_{\iota-\mi \infty }^{\iota+\mi \infty}\frac{\dif s}{2\pi \mi}~ \widetilde{F}^{0}(s) q^{s-3},
	\end{equation}
	with $3<\iota<5$; and
	\begin{equation}
		F^{0}(\lambda) = \int_{\iota-\mi \infty }^{\iota+\mi \infty}\frac{\dif s}{2\pi \mi}~ \widetilde{F}^{0}(s) J(s) \lambda^{-s},
	\end{equation}
	with $3<\iota<5$, where $F_q^{0} = \frac{2}{\pi}\int_{0}^{\infty} \dif \lambda~ \lambda^2 j_1(q \lambda) F^{0}(\lambda) =\frac{1}{8}\frac{q^2}{1-q^2} $ and $J(s) = \int_0^\infty \dif \lambda~ j_1(\lambda) \lambda^{s-1} = \frac{\Gamma(s)}{2-s} \sin\frac{\pi s}{2}$ with $\mre (s)>-1$. 
	Further, we obtain
	\begin{align}
		\hat{\mathbf{A}}_0 F^{0}(\lambda) &= \Big( \frac{16}{\sqrt{3}\lambda^6} +\frac{2}{\sqrt{3}\lambda^4} \Big) \nn\\
		& +  \int_{\iota-\mi \infty }^{\iota+\mi \infty}\frac{\dif s}{2\pi \mi}~ \widetilde{F}^{0}(s) J(s) A_0(s)  \lambda^{-s-3}, 	
	\end{align}
	with $-1<\iota<1$.
	Here we have shifted the integral contour to the left $\iota \to \iota-4$. 
	By doing so, we limit the integral contour in the range $-1<\iota<1$ and then we can work in the momentum space, where the inverse operator $(1-\hat{\mathbf{K}}_1)^{-1}$ is simply $(1-q^2)^{-1}$. 
	For the first part of the right hand side, we find
	\begin{align}
		\label{F1term1}
		\Big(1-\hat{\mathbf{K}}_1\Big)^{-1} & \Big(\frac{16}{\sqrt{3}\lambda^6} +\frac{2}{\sqrt{3}\lambda^4} \Big) = -\frac{2}{15\sqrt{3}} \Big[ \mathrm{Si}(\lambda) j_1(\lambda) \nn\\
		& \qquad\qquad - \mathrm{Ci}(\lambda) y_1(\lambda) -\frac{1}{\lambda^2} - \frac{12}{ \lambda^{4}}  \Big].
	\end{align}
	For the second part, we seek the explicit expression of $(1-\hat{\mathbf{K}}_1)^{-1} \int_{\iota-\mi \infty }^{\iota+\mi \infty}\frac{\dif s}{2\pi \mi}~ \widetilde{F}^{0}(s) J(s) A_0(s)  \lambda^{-s-3}$ in the momentum space and we find  (taking $-1<\iota<1$) 
	\begin{align}
		\frac{1}{1-q^2} \int_{\iota-\mi \infty }^{\iota+\mi \infty}\frac{\dif s}{2\pi \mi}~ \widetilde{F}^{0}(s) J(s) A_0(s) \frac{2}{\pi}J(-s)q^s \nn\\ = -\frac{1}{8}\frac{q}{(1-q^2)^2} + \widetilde{\mathcal{S}}(q).
	\end{align}
	The contour integral becomes an equivalent Fourier transform after a change of variable $s =\iota+ \mi \beta$ and $\me^{\zeta } = q$.
	Here we set $\iota=0$.
	The integrand has poles at $\beta = \pm(2n+1)\mi$, with $n=0,1,2,3,\dots$. 
	For positive or negative values of $\zeta$, we enclose the contour on the upper or lower complex plane, respectively, and sum up all the residues at the poles. 
	Note that there are three general expressions of the residues at $\beta = \pm(6n+1)\mi$, $\pm(6n+3)\mi$, or $\pm(6n+5)\mi$. 
	We find
	\begin{align}
		\widetilde{\mathcal{S}}(q)=& \frac{1}{1-q^2} \Bigg\{ \frac{1}{\pi \sqrt{3} } \Big(q^{2}+q^{-2}+1\Big) \Big[ \Li_2(\me^{\mi \pi/3}q) \nn\\
		&  + \Li_2(\me^{-\mi \pi/3}q) -\Li_2(\me^{2\mi \pi/3}q) -\Li_2(\me^{-2\mi \pi/3}q) \nn\\
		& \mkern-36mu + \frac{\pi}{3}\arctan\Big(\frac{1-q^2}{\sqrt{3}~ q}\Big)  -\frac{\pi^2}{6}  -\ln\Big(\frac{1+q+q^2}{1-q+q^2}\Big) \ln q \Big] \nn\\
		& \mkern-36mu +\Big(-\frac{1}{3} + \frac{2}{\pi\sqrt{3}} \Big) \Big( q - q^{-1} \Big) + \frac{2}{\pi\sqrt{3}}\Big(q +q^{-1}\Big)\ln q \Bigg\},
	\end{align}
	where $\Li_s(z) = \sum_{n=1}^{\infty} \frac{z^n}{n^s}$ is the polylogarithmic function and $\widetilde{\mathcal{S}}(q)$ is a real smooth function of $q$ at $0<q<\infty$.
	Then we obtain a closed-form expression for the first-order solution 
	\begin{align}
		F^{1}(\lambda) 
		&=  -\frac{2}{15\sqrt{3}} \Big[ \mathrm{Si}(\lambda) j_1(\lambda)- \mathrm{Ci}(\lambda) y_1(\lambda) -\frac{1}{\lambda^2} - \frac{12}{ \lambda^{4}}  \Big] \nn\\ 
		& -\frac{\pi}{32}\cos\lambda + \mathcal{S}(\lambda),
	\end{align}
	where $\mathcal{S}(\lambda) = \int_0^\infty \dif q~ q^2 j_1(q\lambda) \widetilde{\mathcal{S}}(q)$ is a smooth and nonoscillatory function. 
	As $\widetilde{\mathcal{S}}(q)  \sim  q \ln q$ at $q\to 0$ and $ \widetilde{\mathcal{S}}(q) \sim q^{-3} \ln q$ at $q\to \infty$, we see $\mathcal{S}(\lambda)  \sim \lambda^{-4} $ at $\lambda\to +\infty$ and $\mathcal{S}(\lambda)  \sim \ln \lambda $ at $\lambda\to 0$.

	Now we seek the perturbative solution to the homogeneous part of \Eq{eqlambda},
	\begin{align}
		\label{eqlambdahomo}
		0= \Big( -1  + \hat{\mathbf{K}}_{1} + \gamma^{3/2}  \hat{\mathbf{A}}_0  \Big) F_{\rm homo}(\lambda). 
	\end{align}
	Up to the first order, the solution takes the form
	\begin{align}
		F_{\rm homo}(\lambda) 
		&= d_{1} \Big[j_1(\lambda) + \gamma^{3/2} \Big(1-\hat{\mathbf{K}}_1\Big)^{-1}\hat{\mathbf{A}}_0 j_1(\lambda)\Big] \nn\\
		& + d_{2} \Big[y_1(\lambda) + \gamma^{3/2} \Big(1-\hat{\mathbf{K}}_1\Big)^{-1}\hat{\mathbf{A}}_0 y_1(\lambda)\Big].
	\end{align}
	Similarly, we write the functions $j_1(\lambda)$ and $y_1(\lambda)$ in the form of the inverse Mellin transforms
	\begin{equation}
		j_1(\lambda)  = \int_{\iota-\mi\infty}^{\iota+\mi\infty}\frac{\dif s}{2\pi \mi}~ J(s)\lambda^{-s}
	\end{equation}
	with $\iota>-1$ and
	\begin{equation}
		y_1(\lambda)  = \int_{\iota-\mi\infty}^{\iota+\mi\infty}\frac{\dif s}{2\pi \mi}~ \Big[-J(s)\cot\frac{\pi s}{2} \Big]\lambda^{-s} 
	\end{equation}
	with $\iota>2$. Then we find 
	\begin{equation}
		\hat{\mathbf{A}}_0 j_1(\lambda) = \int_{\iota-\mi\infty}^{\iota+\mi\infty}\frac{\dif s}{2\pi \mi}~ J(s) A_0(s)\lambda^{-s-3}
	\end{equation}
	with $\iota>-1$ and
	\begin{align}
		\label{A0y1}
		\hat{\mathbf{A}}_0 y_1(\lambda) 
		&=\Big(\frac{128}{3\sqrt{3}} - \frac{48}{\pi}\Big) \lambda^{-5} \nn\\ &+\int_{\iota-\mi\infty}^{\iota+\mi\infty}\frac{\dif s}{2\pi \mi}~ \Big[-J(s) A_0(s) \cot\frac{\pi s}{2} \Big] \lambda^{-s-3}
	\end{align}
	with $0<\iota<2$.
	In \Eq{A0y1} we have shifted the integral contour to the left $\iota \to \iota -2$.
	By doing so, we limit the integral contour in the range $-1<\iota<1$ and then work in the momentum space.
	We find
	\begin{align}
		\frac{2}{\pi}\int_0^\infty \!\!  &\dif \lambda~\lambda^2 j_1(q\lambda) \Big[\hat{\mathbf{A}}_0 j_1(\lambda) \Big] \nn\\ & =\int_{\iota-\mi\infty}^{\iota+\mi\infty}\frac{\dif s}{2\pi \mi}~ J(s) A_0(s) \frac{2}{\pi}J(-s)q^{s} 
	\end{align}
	with $-1<\iota<1$ and
	\begin{align}
		\frac{2}{\pi}&\int_0^\infty \!\! \dif \lambda~\lambda^2 j_1(q\lambda) \Big[\hat{\mathbf{A}}_0 y_1(\lambda) \Big] \nn\\
		&= \Big(-\frac{16}{3\sqrt{3} } +\frac{6}{\pi} \Big) q^2 \nn\\
		&+ \int_{\iota-\mi\infty}^{\iota+\mi\infty}\frac{\dif s}{2\pi \mi}~ \Big[-J(s) A_0(s) \cot\frac{\pi s}{2} \Big] \frac{2}{\pi}J(-s)q^{s}	
	\end{align}
	with $0<\iota<1$.	
	The contour integrals on the right-hand sides can be evaluated using the residue theorem through a large contour on the left or right half complex plane.  
	Then we transform back to the real space and obtain the expressions
	\begin{align}
		(1-\hat{\mathbf{K}}_1)^{-1} &\hat{\mathbf{A}}_0 j_1(\lambda) = \frac{1}{2}\sin\lambda \nn\\
		& \mkern-36mu + \Big(-\frac{16}{\sqrt{3}} + 4\sqrt{3} \ln3\Big) y_1(\lambda) + \mathcal{S}_1(\lambda), \\
		(1-\hat{\mathbf{K}}_1)^{-1} & \hat{\mathbf{A}}_0 y_1(\lambda) =  - \frac{1}{2}\cos\lambda \nn\\
		& \mkern-36mu + \Big(-\frac{16}{3\sqrt{3} } +\frac{6}{\pi} \Big) \Big[ \mathrm{Si}(\lambda) y_1(\lambda)+ \mathrm{Ci}(\lambda) j_1(\lambda) -\frac{2}{\lambda^3}\Big] \nn\\
		&\mkern-36mu +\Big(\frac{9}{2} -\frac{4\pi}{3\sqrt{3}}\Big) y_1(\lambda) + \mathcal{S}_2(\lambda), 
	\end{align}
	where $\mathcal{S}_{1,2}(\lambda) = \int_0^\infty \dif q~ q^2 j_1(q\lambda) \widetilde{\mathcal{S}}_{1,2}(q)$ are  smooth and nonoscillatory functions. 
	Here $\widetilde{\mathcal{S}}_1(q)$ and $\widetilde{\mathcal{S}}_2(q)$ are smooth at $0<q<+\infty$, 
	\begin{align}
		\widetilde{\mathcal{S}}_1(q) &= \frac{1}{1-q^2} \Big[-\frac{16}{\sqrt{3} \pi}\Big(q +q^{-1}\Big) \nn\\
		&\qquad +\frac{8}{\sqrt{3} \pi} \Big(q^2 +q^{-2} + 1\Big) \ln\Big(\frac{1+q+q^{2}}{1-q+q^{2}}\Big) \nn\\ 
		&\qquad -\Big(-\frac{16}{\sqrt{3}} + 4\sqrt{3} \ln3\Big) \frac{2}{\pi} q \Big], \\
		\widetilde{\mathcal{S}}_2(q) &= \frac{1}{1-q^2} \Big[ \frac{8}{\pi} +\Big(\frac{8}{3\sqrt{3}} -\frac{9}{\pi}\Big) q +\frac{16}{3\sqrt{3}} q^2 +\frac{2q}{\pi(1+q)} \nn\\
		&\qquad -\frac{16}{\sqrt{3}\pi} \Big( q^2+q^{-2} +1\Big)\arctan \Big(\frac{\sqrt{3}q^2}{2+q^2}\Big)\Big].
	\end{align} 
	Then we find the asymptotic behaviors of the first-order terms 
	\begin{align}
		&(1-\hat{\mathbf{K}}_1)^{-1} \hat{\mathbf{A}}_0 j_1(\lambda) =\frac{8}{3\sqrt{3}} +\frac{\lambda}{2} + O\Big(\lambda^2\Big) \text{~at~} \lambda\to 0 \\
		&\approx \frac{1}{2}\sin\lambda + \Big(-\frac{16}{\sqrt{3}} + 4\sqrt{3} \ln3\Big) y_1(\lambda) \text{~at~} \lambda\to \infty, \\
		&(1-\hat{\mathbf{K}}_1)^{-1} \hat{\mathbf{A}}_0 y_1(\lambda) = \Big(\frac{32}{3\sqrt{3}} - \frac{12}{\pi} \Big)  \lambda^{-3} \nn\\
		&+ \Big(\frac{32}{3\sqrt{3}} - \frac{24}{\pi} \Big)  \lambda^{-1} + O(\lambda\ln\lambda) \text{~at~} \lambda\to 0 \\
		&= -\frac{1}{2} \cos \lambda + \Big(\frac{15}{2} - \frac{4\pi}{\sqrt{3}} \Big) y_1(\lambda)+ O\Big(\lambda^{-5}\Big) \text{~at~} \lambda\to \infty.
	\end{align}

	The final perturbative solution up to the first order can be expressed as $F(\lambda) = F_{\rm homo}(\lambda) + F^{0}(\lambda) + \gamma^{3/2} F^{1}(\lambda)$. 
	Through the boundary conditions at small and large distances, we can fix the coefficients $d_1$ and $d_2$, and then find the flux of the atom-dimer motion.

	\section{Small $\lambda$ behavior of $F(\lambda)$}
	\label{sect:smalllambda}

	Let us study the small-$\lambda$ behavior of $F(\lambda)$ in \Eq{eqlambda}. 
	Consider the region where $\gamma^{3/2} \simeq \lambda \ll 1 $ or equivalently $R\simeq y \ll l_{\rm d}$. 
	We only retain the two important terms and the homogeneous part of \Eq{eqlambda} becomes 
	\begin{equation}
		0= \Big( \hat{\mathbf{K}}_{1} + \gamma^{3/2}  \hat{\mathbf{A}}_0  \Big) F(\lambda).
	\end{equation}
	We write $F(\lambda)$ in the form of inverse Mellin transform
	\begin{equation}
		\label{S_inverseMellin}
		F(\lambda) = \int_{\iota-\mi \infty}^{\iota+ \mi \infty} \frac{\dif s}{2\pi \mi}~ \tilde{F}(s) \lambda^{-s},
	\end{equation}
	and the equation becomes
	\begin{align}
		\int_{\iota-\mi \infty}^{\iota+ \mi \infty} \frac{\dif s}{2\pi \mi}~ (s-2)(s+1)\tilde{F}(s) \lambda^{-s-2} \nn\\
		= \int_{\iota-\mi \infty}^{\iota+ \mi \infty} \frac{\dif s}{2\pi \mi}~ \gamma^{3/2} A_0(s) \tilde{F}(s) \lambda^{-s-3}. 
	\end{align}
	On the left hand side, we do a contour shift $\iota\to \iota+1$ assuming no poles in the region $\iota< \mre(s) < \iota+1$ and then a change of variable $s \to s+1$. 
	We find that $\tilde{F}(s)$ satisfies
	\begin{equation}
		\label{S_Fs}
		\tilde{F}(s+1) = \frac{\gamma^{3/2} }{\tan \frac{\pi s}{2} } \frac{s^3 G(s)}{ (s-1)(s+2) }  \tilde{F}(s),
	\end{equation}
	where $A_0(s) \equiv \frac{s^3}{\tan \frac{\pi s}{2} } G(s)$ and
	\begin{equation}
		G(s)  =  1 - \frac{4}{s^2} \Big(1+8 \frac{\cos \frac{\pi s}{6}}{\cos \frac{\pi s}{2}}\Big) + \frac{64}{\sqrt{3} s^3} \frac{\sin \frac{\pi s}{6}}{\cos \frac{\pi s}{2}} .
	\end{equation}
	The function $G(s)$ has zeros at $\pm u_0$, $\pm u_0^\ast$, and $\pm u_n$ with $n =1, 2, 3, \dots$ and poles at $0$ and $\pm b_n$ with $n=0,1,2,3\dots$, where 
	\begin{align}
		&u_0 \approx 2.086+ 1.103 \mi,\quad u_1= 4,\quad u_2 = 6, \nn\\
		&u_3\approx 6.39,\quad u_4\approx 8.97,\quad u_5\approx 11.16, \dots, \\
		&b_n = 2n+1, 
	\end{align}
	and at $n\to +\infty$, 
	\begin{align}
		&u_n-b_n \nn\\
		&= \begin{cases}\tiny
			\!\!-\frac{8}{3\sqrt{3}\pi} \frac{1}{n^2} + \frac{32}{27\sqrt{3} \pi} \frac{1}{n^3} + O\Big(\frac{1}{n^4}\Big)  
			&\!\!\!\!\!\text{if}\!\!\!\! \mod\!\!(2n+1,6)\!=\!1 \\
			\!\!-\frac{16}{27\sqrt{3} \pi} \frac{1}{n^3} + O\Big(\frac{1}{n^4}\Big) 
			&\!\!\!\!\!\text{if}\!\!\!\! \mod\!\!(2n+1,6)\!=\!3 \\
			\!\!\frac{8}{3\sqrt{3}\pi} \frac{1}{n^2} - \frac{112}{27\sqrt{3} \pi} \frac{1}{n^3} + O\Big(\frac{1}{n^4}\Big) 
			&\!\!\!\!\!\text{if}\!\!\!\! \mod\!\!(2n+1,6)\!=\!5 .
		\end{cases}
	\end{align}
	Then $G(s)$ can be expressed in the infinite product form
	\begin{equation}
		G(s) =\frac{(s^2-u_0^2)(s^2-u_0^{\ast 2})}{s^2 (s^2-1)} \prod_{n=1}^{+\infty}\frac{s^2-u_n^2}{s^2-b_n^2}.
	\end{equation}

	The correct solution of $\tilde{F}(s)$ should (i) satisfy \Eq{S_Fs}, (ii) be integrable in \Eq{S_inverseMellin}, and (iii) have no singularities on the certain stripe on the complex plane. 
	Then we construct one solution
	\begin{align}
		\tilde{F}_1(s) &= \gamma^{3s/2} \sin\Big(\frac{\pi s}{2}\Big) \frac{\Gamma(1-s+1)}{\Gamma(s+2)} \Gamma(s)  \frac{\Gamma(s+u_0^\ast)}{\Gamma(1-s+u_0^\ast)} \nn\\
		&\qquad \times \prod_{n=0}^{+\infty} \frac{\Gamma(s+u_n) \Gamma(1-s+b_n)}{\Gamma(s+b_n) \Gamma(1-s+u_n)}.
	\end{align}
	Here $\tilde{F}_1(s)$ has no pole on the strip $-\mre(u_0) < \mre(s) < 2$. The closest poles to the imaginary axis are $s = -u_0$ and $s =-u_0^{\ast}$.
	Thus we see that at small $\lambda$,
	\begin{align}
		F_1(\lambda) &\sim c_1 \Big( \frac{\lambda}{\gamma^{3/2}} \Big)^{u_0} + c_1^\ast \Big(\frac{\lambda}{\gamma^{3/2}} \Big)^{u_0^\ast} \nn\\
		&\approx 2|c_1| \Big( \frac{\lambda}{\gamma^{3/2}} \Big)^{2.086} \sin\Big(1.103 \ln \frac{\lambda}{\gamma^{3/2}} +\theta_1 +\frac{\pi}{2} \Big),
	\end{align}
	where $c_1 = |c_1|\me^{\mi \theta_1}$ is a complex number independent of $\gamma$ and $\lambda$. 
	Note that $\lambda /\gamma^{3/2} = y/R$. 
	The other solution $F_2(\lambda)$ will be constructed in a similar way and have a similar small-$\lambda$ behavior (with a different phase inside the sine function). 
	Therefore, at small $\lambda$, the solution to the homogeneous part of \Eq{eqlambda} has the behavior
	\begin{equation}
		F(\lambda) \propto \Big( \frac{\lambda}{\gamma^{3/2}} \Big)^{2.086} \sin\Big(1.103 \ln \frac{\lambda}{\gamma^{3/2}} +\overbar{\theta} +\frac{\pi}{2} \Big).
	\end{equation}

	\section{Thermal average}
	\label{sect:ta}
	
	Consider the thermal average of the three-body recombination rate constant. 
	Based on the argument of Ref.~\cite{Burke1999}, we consider the distribution function $f(\vect{r}_1, \vect{r}_2, \vect{r}_3, \overbar{\vect{b}}_1, \overbar{\vect{b}}_2, \overbar{\vect{b}}_3)$ for the three atoms with position vectors $\vect{r}_1, \vect{r}_2, \vect{r}_3$ and momenta $\overbar{\vect{b}}_1, \overbar{\vect{b}}_2, \overbar{\vect{b}}_3$. 
	The distribution function is normalized as 
	\begin{equation}
		\int \dif^3 r_1\dif^3 r_2\dif^3 r_3 \dif^3 \overbar{b}_1\dif^3 \overbar{b}_2\dif^3 \overbar{b}_3 ~f(\vect{r}_1, \vect{r}_2, \vect{r}_3, \overbar{\vect{b}}_1, \overbar{\vect{b}}_2, \overbar{\vect{b}}_3) = 1 .
	\end{equation}
	Then, the thermal average of the flux $\Phi$ of the atom-dimer outgoing wave is
	\begin{align}
		\langle \Phi \rangle_T &= \int \dif^3 r_1\dif^3 r_2\dif^3 r_3 \dif^3 \overbar{b}_1\dif^3 \overbar{b}_2\dif^3 \overbar{b}_3 \nn\\
		&\qquad \times f(\vect{r}_1, \vect{r}_2, \vect{r}_3, \overbar{\vect{b}}_1, \overbar{\vect{b}}_2, \overbar{\vect{b}}_3) \Phi .
	\end{align}
	Here $\Phi$ is a function of the momenta $(\vect{b},\vect{B})$, $\Phi = \Phi(\vect{b}, \vect{B})$. 
	According to the Maxwell-Boltzmann distribution, $f(\vect{r}_1, \vect{r}_2, \vect{r}_3, \overbar{\vect{b}}_1, \overbar{\vect{b}}_2, \overbar{\vect{b}}_3) \propto \exp(-H/k_B T)$. 
	For a homogeneous gas, neglecting the interactions, the Hamiltonian
	\begin{equation}
		H = \frac{1}{2}\Big(\overbar{\vect{b}}_1^{2} + \overbar{\vect{b}}_2^{2} +\overbar{\vect{b}}_3^{2} \Big) =\frac{1}{6}b_c^2 + b^2 +B^2.
	\end{equation} 
	Then, 
	\begin{equation}
		\langle \Phi(\vect{b}, \vect{B}) \rangle_T = \frac{\int \dif^3 b \dif^3 B ~    \me^{-(b^2+B^2)/k_B T} \Phi(\vect{b}, \vect{B}) }{ \int \dif^3 b \dif^3 B ~    \me^{-(b^2+B^2)/k_B T}  } .
	\end{equation}
	
	If $\Phi(\vect{b}, \vect{B}) = \Phi(E)$ with $E = b^2+B^2$, then
	\begin{equation}
		\langle \Phi(E) \rangle_T = \frac{1}{2(k_B T)^3} \int_0^\infty \dif E~ E^2 \Phi(E) \me^{-E/k_BT}.
	\end{equation}
	This recovers Eq.~(3) of Ref.~\cite{Yoshida2018}.

	Note that if the integrand only depends on $b$, $B$, and the angle $\theta$ between $\vect{b}$ and $\vect{B}$, we can use $\int\dif^3 b \dif^3 B = \int_{0}^{\infty} \dif b~ 4\pi b^2 \int_{0}^{\infty} \dif B~ 2\pi B^2 \int_{0}^{\pi}\dif \theta~ \sin\theta$ to evaluate the formula. 
	Here we list the formulas
	\begin{align}
		& \langle |\vect{b}\times\vect{B}|^2  \rangle_T = \frac{3}{2} (k_BT)^2, \\
		& \langle b^2 (b^2-B^2)^2 + 4 b^2 B^2 (2B^2 - b^2) t^2 \rangle_T = 15 (k_BT)^3, \\
		& \langle b^6 +\frac{2}{115} b^4 B^2 (208-673 t^2) \nn\\
		&\qquad\qquad + \frac{1}{155} b^2 B^4 (881 + 514 t^2)\rangle_T = \frac{1554}{31} (k_BT)^3, 
	\end{align}
	where $t \equiv \hat{\vect{b}}\cdot \hat{\vect{B}} \equiv \cos\theta$.


\begin{thebibliography}{41}
		\expandafter\ifx\csname natexlab\endcsname\relax\def\natexlab#1{#1}\fi
		\expandafter\ifx\csname bibnamefont\endcsname\relax
		\def\bibnamefont#1{#1}\fi
		\expandafter\ifx\csname bibfnamefont\endcsname\relax
		\def\bibfnamefont#1{#1}\fi
		\expandafter\ifx\csname citenamefont\endcsname\relax
		\def\citenamefont#1{#1}\fi
		\expandafter\ifx\csname url\endcsname\relax
		\def\url#1{\texttt{#1}}\fi
		\expandafter\ifx\csname urlprefix\endcsname\relax\def\urlprefix{URL }\fi
		\providecommand{\bibinfo}[2]{#2}
		\providecommand{\eprint}[2][]{\url{#2}}
		
		\bibitem{Esry1999} B. D. Esry, C. H. Greene, and J. P. Burke, Phys. Rev. Lett. \textbf{83}, 1751 (1999).
		
		\bibitem{Kraemer2006} T. Kraemer, M. Mark, P. Waldburger, J. G. Danzl, C. Chin, B. Engeser, A. D. Lange, K. Pilch, A. Jaakkola, H.-C. Nägerl and R. Grimm, Nature \textbf{440}, 315 (2006).
		
		\bibitem{Hammer2007} H.-W. Hammer and L. Platter, Eur. Phys. J. A \textbf{32}, 113 (2007).
		
		\bibitem{Stecher2009} J. von Stecher, J. P. D’Incao, and C. H. Greene, Nat. Phys. \textbf{5}, 417 (2009).
		
		\bibitem{Ferlaino2009} F. Ferlaino, S. Knoop, M. Berninger, W. Harm, J. P. D’Incao, H.-C. Nägerl, and R. Grimm, Phys. Rev. Lett. \textbf{102}, 140401 (2009).	
		
		\bibitem{Efimov2009a} S. Knoop, F. Ferlaino, M. Mark, M. Berninger, H. Sch\"{o}bel, H.-C. N\"{a}gerl and R. Grimm, Nat. Phys. {\bf 5} 227 (2009)
		
		\bibitem{Efimov2009b} M. Zaccanti, B. Deissler, C. D’Errico, M. Fattori, M. Jona-Lasinio, S. M\"{u}ller, G. Roati, M. Inguscio and G. Modugno, Nat. Phys. {\bf 5}, 586 (2009) 
		
		\bibitem{Efimov2017} J. Johansen, B. J. DeSalvo, K. Patel and C. Chin, Nat. Phys. {\bf 13} 731 (2017). 
		
		\bibitem{Efimov2020} X. Xie, M. J. Van de Graaff, R. Chapurin, M. D. Frye, J. M. Hutson, J. P. D’Incao, P. S. Julienne, J. Ye, and E. A. Cornell, Phys. Rev. Lett. {\bf 125}, 243401 (2020) 		
		
		\bibitem[{\citenamefont{Levinsen et~al.}(2007)\citenamefont{Levinsen, Cooper,
				and Gurarie}}]{Levinsen2007}
		\bibinfo{author}{\bibfnamefont{J.}~\bibnamefont{Levinsen}},
		\bibinfo{author}{\bibfnamefont{N.~R.} \bibnamefont{Cooper}},
		\bibnamefont{and} \bibinfo{author}{\bibfnamefont{V.}~\bibnamefont{Gurarie}},
		\bibinfo{journal}{Phys. Rev. Lett.} \textbf{\bibinfo{volume}{99}},
		\bibinfo{pages}{210402} (\bibinfo{year}{2007}).
		
		\bibitem[{\citenamefont{Levinsen et~al.}(2008)\citenamefont{Levinsen, Cooper,
				and Gurarie}}]{Levinsen2008}
		\bibinfo{author}{\bibfnamefont{J.}~\bibnamefont{Levinsen}},
		\bibinfo{author}{\bibfnamefont{N.~R.} \bibnamefont{Cooper}},
		\bibnamefont{and} \bibinfo{author}{\bibfnamefont{V.}~\bibnamefont{Gurarie}},
		\bibinfo{journal}{Phys. Rev. A} \textbf{\bibinfo{volume}{78}},
		\bibinfo{pages}{063616} (\bibinfo{year}{2008}).
		
		\bibitem[{\citenamefont{Cooper and Shlyapnikov}(2009)}]{Cooper2009}
		\bibinfo{author}{\bibfnamefont{N.~R.} \bibnamefont{Cooper}} \bibnamefont{and}
		\bibinfo{author}{\bibfnamefont{G.~V.} \bibnamefont{Shlyapnikov}},
		\bibinfo{journal}{Phys. Rev. Lett.} \textbf{\bibinfo{volume}{103}},
		\bibinfo{pages}{155302} (\bibinfo{year}{2009}).
		
		\bibitem[{\citenamefont{Han et~al.}(2009)\citenamefont{Han, Chan, Yi, Daley,
				Diehl, Zoller, and Duan}}]{Han2009}
		\bibinfo{author}{\bibfnamefont{Y.-J.} \bibnamefont{Han}},
		\bibinfo{author}{\bibfnamefont{Y.-H.} \bibnamefont{Chan}},
		\bibinfo{author}{\bibfnamefont{W.}~\bibnamefont{Yi}},
		\bibinfo{author}{\bibfnamefont{A.~J.} \bibnamefont{Daley}},
		\bibinfo{author}{\bibfnamefont{S.}~\bibnamefont{Diehl}},
		\bibinfo{author}{\bibfnamefont{P.}~\bibnamefont{Zoller}}, \bibnamefont{and}
		\bibinfo{author}{\bibfnamefont{L.-M.} \bibnamefont{Duan}},
		\bibinfo{journal}{Phys. Rev. Lett.} \textbf{\bibinfo{volume}{103}},
		\bibinfo{pages}{070404} (\bibinfo{year}{2009}).
		
		\bibitem[{\citenamefont{Zinner}(2010)}]{Zinner2010}
		\bibinfo{author}{\bibfnamefont{N.~T.} \bibnamefont{Zinner}},
		\bibinfo{journal}{Eur. Phys. J. D} \textbf{\bibinfo{volume}{57}},
		\bibinfo{pages}{235} (\bibinfo{year}{2010}).
		
		\bibitem[{\citenamefont{Waseem et~al.}(2017)\citenamefont{Waseem, Saito,
				Yoshida, and Mukaiyama}}]{Waseem2017}
		\bibinfo{author}{\bibfnamefont{M.}~\bibnamefont{Waseem}},
		\bibinfo{author}{\bibfnamefont{T.}~\bibnamefont{Saito}},
		\bibinfo{author}{\bibfnamefont{J.}~\bibnamefont{Yoshida}}, \bibnamefont{and}
		\bibinfo{author}{\bibfnamefont{T.}~\bibnamefont{Mukaiyama}},
		\bibinfo{journal}{Phys. Rev. A} \textbf{\bibinfo{volume}{96}},
		\bibinfo{pages}{062704} (\bibinfo{year}{2017}).
		
		\bibitem[{\citenamefont{Kurlov and Shlyapnikov}(2017)}]{Kurlov2017}
		\bibinfo{author}{\bibfnamefont{D.~V.} \bibnamefont{Kurlov}} \bibnamefont{and}
		\bibinfo{author}{\bibfnamefont{G.~V.} \bibnamefont{Shlyapnikov}},
		\bibinfo{journal}{Phys. Rev. A} \textbf{\bibinfo{volume}{95}},
		\bibinfo{pages}{032710} (\bibinfo{year}{2017}).
		
		\bibitem[{\citenamefont{Zhou and Cui}(2017)}]{Zhou2017}
		\bibinfo{author}{\bibfnamefont{L.}~\bibnamefont{Zhou}} \bibnamefont{and}
		\bibinfo{author}{\bibfnamefont{X.}~\bibnamefont{Cui}},
		\bibinfo{journal}{Phys. Rev. A} \textbf{\bibinfo{volume}{96}},
		\bibinfo{pages}{030701} (\bibinfo{year}{2017}).
		
		\bibitem[{\citenamefont{Pan et~al.}(2018)\citenamefont{Pan, Chen, and
				Cui}}]{Pan2018}
		\bibinfo{author}{\bibfnamefont{L.}~\bibnamefont{Pan}},
		\bibinfo{author}{\bibfnamefont{S.}~\bibnamefont{Chen}}, \bibnamefont{and}
		\bibinfo{author}{\bibfnamefont{X.}~\bibnamefont{Cui}},
		\bibinfo{journal}{Phys. Rev. A} \textbf{\bibinfo{volume}{98}},
		\bibinfo{pages}{011603} (\bibinfo{year}{2018}).
		
		\bibitem[{\citenamefont{Hu et~al.}(2018)\citenamefont{Hu, Mulkerin, He, Wang,
				and Liu}}]{Hu2018}
		\bibinfo{author}{\bibfnamefont{H.}~\bibnamefont{Hu}},
		\bibinfo{author}{\bibfnamefont{B.~C.} \bibnamefont{Mulkerin}},
		\bibinfo{author}{\bibfnamefont{L.}~\bibnamefont{He}},
		\bibinfo{author}{\bibfnamefont{J.}~\bibnamefont{Wang}}, \bibnamefont{and}
		\bibinfo{author}{\bibfnamefont{X.-J.} \bibnamefont{Liu}},
		\bibinfo{journal}{Phys. Rev. A} \textbf{\bibinfo{volume}{98}},
		\bibinfo{pages}{063605} (\bibinfo{year}{2018}).
		
		\bibitem[{\citenamefont{Jiang and Zhou}(2018)}]{Jiang2018}
		\bibinfo{author}{\bibfnamefont{S.-J.} \bibnamefont{Jiang}} \bibnamefont{and}
		\bibinfo{author}{\bibfnamefont{F.}~\bibnamefont{Zhou}},
		\bibinfo{journal}{Phys. Rev. A} \textbf{\bibinfo{volume}{97}},
		\bibinfo{pages}{063606} (\bibinfo{year}{2018}).
		
		\bibitem[{\citenamefont{Chang et~al.}(2020)\citenamefont{Chang, Senaratne,
				Cavazos-Cavazos, and Hulet}}]{Chang2020}
		\bibinfo{author}{\bibfnamefont{Y.-T.} \bibnamefont{Chang}},
		\bibinfo{author}{\bibfnamefont{R.}~\bibnamefont{Senaratne}},
		\bibinfo{author}{\bibfnamefont{D.}~\bibnamefont{Cavazos-Cavazos}},
		\bibnamefont{and} \bibinfo{author}{\bibfnamefont{R.~G.} \bibnamefont{Hulet}},
		\bibinfo{journal}{Phys. Rev. Lett.} \textbf{\bibinfo{volume}{125}},
		\bibinfo{pages}{263402} (\bibinfo{year}{2020}).
		
		\bibitem[{\citenamefont{Tan}(2008)}]{Tan2008}
		\bibinfo{author}{\bibfnamefont{S.}~\bibnamefont{Tan}}, \bibinfo{journal}{Phys.
			Rev. A} \textbf{\bibinfo{volume}{78}}, \bibinfo{pages}{013636}
		(\bibinfo{year}{2008}).
		
		\bibitem[{Zhu()}]{Zhu2017}
		\bibinfo{note}{S. Zhu and S. Tan, arXiv:1710.04147.}
		
		\bibitem[{\citenamefont{Mestrom et~al.}(2019)\citenamefont{Mestrom, Colussi,
				Secker, and Kokkelmans}}]{Mestrom2019}
		\bibinfo{author}{\bibfnamefont{P.~M.~A.} \bibnamefont{Mestrom}},
		\bibinfo{author}{\bibfnamefont{V.~E.} \bibnamefont{Colussi}},
		\bibinfo{author}{\bibfnamefont{T.}~\bibnamefont{Secker}}, \bibnamefont{and}
		\bibinfo{author}{\bibfnamefont{S.~J. J. M.~F.} \bibnamefont{Kokkelmans}},
		\bibinfo{journal}{Phys. Rev. A} \textbf{\bibinfo{volume}{100}},
		\bibinfo{pages}{050702} (\bibinfo{year}{2019}).
		
		\bibitem[{\citenamefont{Wang and Tan}(2021)}]{Wang2021}
		\bibinfo{author}{\bibfnamefont{Z.}~\bibnamefont{Wang}} \bibnamefont{and}
		\bibinfo{author}{\bibfnamefont{S.}~\bibnamefont{Tan}},
		\bibinfo{journal}{Phys. Rev. A} \textbf{\bibinfo{volume}{104}},
		\bibinfo{pages}{043319} (\bibinfo{year}{2021}).
		
		\bibitem[{\citenamefont{Esry et~al.}(2001)\citenamefont{Esry, Greene, and
				Suno}}]{Esry2001}
		\bibinfo{author}{\bibfnamefont{B.~D.} \bibnamefont{Esry}},
		\bibinfo{author}{\bibfnamefont{C.~H.} \bibnamefont{Greene}},
		\bibnamefont{and} \bibinfo{author}{\bibfnamefont{H.}~\bibnamefont{Suno}},
		\bibinfo{journal}{Phys. Rev. A} \textbf{\bibinfo{volume}{65}},
		\bibinfo{pages}{010705} (\bibinfo{year}{2001}).
		
		\bibitem[{\citenamefont{Suno et~al.}(2003)\citenamefont{Suno, Esry, and
				Greene}}]{Suno2003}
		\bibinfo{author}{\bibfnamefont{H.}~\bibnamefont{Suno}},
		\bibinfo{author}{\bibfnamefont{B.~D.} \bibnamefont{Esry}}, \bibnamefont{and}
		\bibinfo{author}{\bibfnamefont{C.~H.} \bibnamefont{Greene}},
		\bibinfo{journal}{Phys. Rev. Lett.} \textbf{\bibinfo{volume}{90}},
		\bibinfo{pages}{053202} (\bibinfo{year}{2003}).
		
		\bibitem[{\citenamefont{Schmidt et~al.}(2020)\citenamefont{Schmidt, Hammer, and
				Platter}}]{Schmidt2020}
		\bibinfo{author}{\bibfnamefont{M.}~\bibnamefont{Schmidt}},
		\bibinfo{author}{\bibfnamefont{H.-W.} \bibnamefont{Hammer}},
		\bibnamefont{and} \bibinfo{author}{\bibfnamefont{L.}~\bibnamefont{Platter}},
		\bibinfo{journal}{Phys. Rev. A} \textbf{\bibinfo{volume}{101}},
		\bibinfo{pages}{062702} (\bibinfo{year}{2020}).
		
		\bibitem[{\citenamefont{Yoshida et~al.}(2018)\citenamefont{Yoshida, Saito,
				Waseem, Hattori, and Mukaiyama}}]{Yoshida2018}
		\bibinfo{author}{\bibfnamefont{J.}~\bibnamefont{Yoshida}},
		\bibinfo{author}{\bibfnamefont{T.}~\bibnamefont{Saito}},
		\bibinfo{author}{\bibfnamefont{M.}~\bibnamefont{Waseem}},
		\bibinfo{author}{\bibfnamefont{K.}~\bibnamefont{Hattori}}, \bibnamefont{and}
		\bibinfo{author}{\bibfnamefont{T.}~\bibnamefont{Mukaiyama}},
		\bibinfo{journal}{Phys. Rev. Lett.} \textbf{\bibinfo{volume}{120}},
		\bibinfo{pages}{133401} (\bibinfo{year}{2018}).
		
		\bibitem[{\citenamefont{Waseem et~al.}(2018)\citenamefont{Waseem, Yoshida,
				Saito, and Mukaiyama}}]{Waseem2018}
		\bibinfo{author}{\bibfnamefont{M.}~\bibnamefont{Waseem}},
		\bibinfo{author}{\bibfnamefont{J.}~\bibnamefont{Yoshida}},
		\bibinfo{author}{\bibfnamefont{T.}~\bibnamefont{Saito}}, \bibnamefont{and}
		\bibinfo{author}{\bibfnamefont{T.}~\bibnamefont{Mukaiyama}},
		\bibinfo{journal}{Phys. Rev. A} \textbf{\bibinfo{volume}{98}},
		\bibinfo{pages}{020702} (\bibinfo{year}{2018}).
		
		\bibitem[{\citenamefont{Waseem et~al.}(2019)\citenamefont{Waseem, Yoshida,
				Saito, and Mukaiyama}}]{Waseem2019}
		\bibinfo{author}{\bibfnamefont{M.}~\bibnamefont{Waseem}},
		\bibinfo{author}{\bibfnamefont{J.}~\bibnamefont{Yoshida}},
		\bibinfo{author}{\bibfnamefont{T.}~\bibnamefont{Saito}}, \bibnamefont{and}
		\bibinfo{author}{\bibfnamefont{T.}~\bibnamefont{Mukaiyama}},
		\bibinfo{journal}{Phys. Rev. A} \textbf{\bibinfo{volume}{99}},
		\bibinfo{pages}{052704} (\bibinfo{year}{2019}).
		
		\bibitem[{\citenamefont{Jona-Lasinio et~al.}(2008)\citenamefont{Jona-Lasinio,
				Pricoupenko, and Castin}}]{Jona-Lasinio2008}
		\bibinfo{author}{\bibfnamefont{M.}~\bibnamefont{Jona-Lasinio}},
		\bibinfo{author}{\bibfnamefont{L.}~\bibnamefont{Pricoupenko}},
		\bibnamefont{and} \bibinfo{author}{\bibfnamefont{Y.}~\bibnamefont{Castin}},
		\bibinfo{journal}{Phys. Rev. A} \textbf{\bibinfo{volume}{77}},
		\bibinfo{pages}{043611} (\bibinfo{year}{2008}).
		
		\bibitem[{\citenamefont{Pricoupenko}(2006)}]{Pricoupenko2006}
		\bibinfo{author}{\bibfnamefont{L.}~\bibnamefont{Pricoupenko}},
		\bibinfo{journal}{Phys. Rev. Lett.} \textbf{\bibinfo{volume}{96}},
		\bibinfo{pages}{050401} (\bibinfo{year}{2006}).
		
		\bibitem[{\citenamefont{Peng et~al.}(2014)\citenamefont{Peng, Tan, and
				Jiang}}]{Peng2014}
		\bibinfo{author}{\bibfnamefont{S.-G.} \bibnamefont{Peng}},
		\bibinfo{author}{\bibfnamefont{S.}~\bibnamefont{Tan}}, \bibnamefont{and}
		\bibinfo{author}{\bibfnamefont{K.}~\bibnamefont{Jiang}},
		\bibinfo{journal}{Phys. Rev. Lett.} \textbf{\bibinfo{volume}{112}},
		\bibinfo{pages}{250401} (\bibinfo{year}{2014}).
		
		
		\bibitem[{\citenamefont{Fox}(1957)}]{Fox1957}
		\bibinfo{author}{\bibfnamefont{C.}~\bibnamefont{Fox}},
		\bibinfo{journal}{Canadian Journal of Mathematics}
		\textbf{\bibinfo{volume}{9}}, \bibinfo{pages}{110–117}
		(\bibinfo{year}{1957}).
		
		\bibitem[{Sko()}]{Skorniakov1956}
		\bibinfo{note}{G. V. Skorniakov and K. A. Ter-Martirosian, Zh. Eksp. Teor. Fiz.
			\textbf{31}, 775 (1956) [Sov. Phys. JETP \textbf{4}, 648 (1957)].}
		
		\bibitem[{\citenamefont{Petrov}(2004)}]{Petrov2004}
		\bibinfo{author}{\bibfnamefont{D.~S.} \bibnamefont{Petrov}},
		\bibinfo{journal}{Phys. Rev. Lett.} \textbf{\bibinfo{volume}{93}},
		\bibinfo{pages}{143201} (\bibinfo{year}{2004}).
		
		\bibitem[{\citenamefont{Braaten et~al.}(2012)\citenamefont{Braaten, Hagen,
				Hammer, and Platter}}]{Braaten2012}
		\bibinfo{author}{\bibfnamefont{E.}~\bibnamefont{Braaten}},
		\bibinfo{author}{\bibfnamefont{P.}~\bibnamefont{Hagen}},
		\bibinfo{author}{\bibfnamefont{H.-W.} \bibnamefont{Hammer}},
		\bibnamefont{and} \bibinfo{author}{\bibfnamefont{L.}~\bibnamefont{Platter}},
		\bibinfo{journal}{Phys. Rev. A} \textbf{\bibinfo{volume}{86}},
		\bibinfo{pages}{012711} (\bibinfo{year}{2012}).
		
		\bibitem[{\citenamefont{Nishida}(2012)}]{Nishida2012}
		\bibinfo{author}{\bibfnamefont{Y.}~\bibnamefont{Nishida}},
		\bibinfo{journal}{Phys. Rev. A} \textbf{\bibinfo{volume}{86}},
		\bibinfo{pages}{012710} (\bibinfo{year}{2012}).
		
		\bibitem{Efimov1970} V. N. Efimov, Yad. Fiz. \textbf{12}, 1080 (1970) [Sov. J. Nucl. Phys. \textbf{12}, 589 (1971)]. 
		
		\bibitem[{\citenamefont{Efimov}(1970{\natexlab{b}})}]{Efimov1970a}
		\bibinfo{author}{\bibfnamefont{V.}~\bibnamefont{Efimov}},
		\bibinfo{journal}{Phys. Lett. B} \textbf{\bibinfo{volume}{33}},
		\bibinfo{pages}{563} (\bibinfo{year}{1970}{\natexlab{b}}).
		
		\bibitem[{\citenamefont{Braaten and Hammer}(2006)}]{Braaten2006}
		\bibinfo{author}{\bibfnamefont{E.}~\bibnamefont{Braaten}} \bibnamefont{and}
		\bibinfo{author}{\bibfnamefont{H.-W.} \bibnamefont{Hammer}},
		\bibinfo{journal}{Physics Reports} \textbf{\bibinfo{volume}{428}},
		\bibinfo{pages}{259} (\bibinfo{year}{2006}).
		
		\bibitem[{\citenamefont{Braaten and Hammer}(2007)}]{Braaten2007}
		\bibinfo{author}{\bibfnamefont{E.}~\bibnamefont{Braaten}} \bibnamefont{and}
		\bibinfo{author}{\bibfnamefont{H.-W.} \bibnamefont{Hammer}},
		\bibinfo{journal}{Annals of Physics} \textbf{\bibinfo{volume}{322}},
		\bibinfo{pages}{120} (\bibinfo{year}{2007}).
		
		\bibitem[{\citenamefont{Naidon and Endo}(2017)}]{Naidon2017}
		\bibinfo{author}{\bibfnamefont{P.}~\bibnamefont{Naidon}} \bibnamefont{and}
		\bibinfo{author}{\bibfnamefont{S.}~\bibnamefont{Endo}},
		\bibinfo{journal}{Rep. Prog. Phys.} \textbf{\bibinfo{volume}{80}},
		\bibinfo{pages}{056001} (\bibinfo{year}{2017}).
		
		\bibitem[{\citenamefont{Nishida and Tan}(2011)}]{Nishida2011}
		\bibinfo{author}{\bibfnamefont{Y.}~\bibnamefont{Nishida}} \bibnamefont{and}
		\bibinfo{author}{\bibfnamefont{S.}~\bibnamefont{Tan}},
		\bibinfo{journal}{Few-Body Syst.} \textbf{\bibinfo{volume}{51}},
		\bibinfo{pages}{191} (\bibinfo{year}{2011}).
		
		\bibitem[{\citenamefont{Zhu and Tan}(2013)}]{Zhu2013}
		\bibinfo{author}{\bibfnamefont{S.}~\bibnamefont{Zhu}} \bibnamefont{and}
		\bibinfo{author}{\bibfnamefont{S.}~\bibnamefont{Tan}},
		\bibinfo{journal}{Phys. Rev. A} \textbf{\bibinfo{volume}{87}},
		\bibinfo{pages}{063629} (\bibinfo{year}{2013}).
		
		\bibitem[{\citenamefont{Efremov et~al.}(2013)\citenamefont{Efremov, Plimak,
				Ivanov, and Schleich}}]{Efremov2013}
		\bibinfo{author}{\bibfnamefont{M.~A.} \bibnamefont{Efremov}},
		\bibinfo{author}{\bibfnamefont{L.}~\bibnamefont{Plimak}},
		\bibinfo{author}{\bibfnamefont{M.~Y.} \bibnamefont{Ivanov}},
		\bibnamefont{and} \bibinfo{author}{\bibfnamefont{W.~P.}
			\bibnamefont{Schleich}}, \bibinfo{journal}{Phys. Rev. Lett.}
		\textbf{\bibinfo{volume}{111}}, \bibinfo{pages}{113201}
		(\bibinfo{year}{2013}).
		
		\bibitem[{\citenamefont{Mora et~al.}(2011)\citenamefont{Mora, Gogolin, and
				Egger}}]{Mora2011}
		\bibinfo{author}{\bibfnamefont{C.}~\bibnamefont{Mora}},
		\bibinfo{author}{\bibfnamefont{A.~O.} \bibnamefont{Gogolin}},
		\bibnamefont{and} \bibinfo{author}{\bibfnamefont{R.}~\bibnamefont{Egger}},
		\bibinfo{journal}{Comptes Rendus Physique} \textbf{\bibinfo{volume}{12}},
		\bibinfo{pages}{27} (\bibinfo{year}{2011}).
		
		\bibitem[{Bur()}]{Burke1999}
		\bibinfo{note}{J. P. Burke, Jr., Ph.D. thesis, University of Colorado, 1999.}
		
	\end{thebibliography}

\end{document}